\documentclass[aps,pra,preprint,preprintnumbers,,showpacs,a4paper,twoside,12pt]{revtex4-1}

\usepackage{amsmath,amssymb}
\usepackage{graphicx}   
\usepackage{color,dsfont}  

\newcommand{\matrice}[2][cccccccccccccccccc]{\left(\hspace{-2mm}\begin{array}{#1}#2\\ \end{array}\hspace{-1mm}\right)}

\let\theta=\vartheta
\let\rho=\varrho
\let\epsilon=\varepsilon
\let\setminus=\backslash

\def\ie{i.e.~}
\def\eg{e.g.~}

\def\real{{\mathbb R}}
\def\0{{}}
\def\eref#1{(\ref{#1})}
\def\Section{{Sec.}}

\def\Fig{{Fig.~}}
\def\Figs{{Figs.~}}

\begin{document}

\title{Escape behaviour of a quantum particle\\ in a loop coupled to a lead}

\author{Ph.~A.~Jacquet}
\email[E-mail: ]{philippe.jacquet@kwansei.ac.jp}

\affiliation{Department of Physics, Kwansei Gakuin University, Sanda 669-1337, Japan}

\received{14 October 2011}

\begin{abstract}
\noindent
We consider a one-dimensional loop of circumference $L$ crossed by a constant magnetic flux $\Phi$ and connected to an infinite lead with coupling parameter $\epsilon$. Assuming that the initial state $\psi_0$ of the particle is confined inside the loop and evolves freely, we analyse the time evolution of the nonescape probability $P(\psi_0,L,\Phi,\epsilon,t)$, which is the probability that the particle will still be inside the loop at some later time $t$. In appropriate units, we found that $P(\psi_0,L,\Phi,\epsilon,t) = P_\infty(\psi_0,\Phi) + \sum_{k=1}^\infty C_k(\psi_0,L,\Phi,\epsilon)/t^k$. The constant $P_\infty(\psi_0,\Phi)$ is independent of $L$ and $\epsilon$, and vanishes if $\psi_0$ has no bound state components or if $|\cos(\Phi)|\not=1$. The coefficients $C_1(\psi_0,L,\Phi,\epsilon)$ and $C_3(\psi_0,L,\Phi,\epsilon)$ depend on the initial state $\psi_0$ of the particle, but only the momentum $k=\Phi/L$ is involved. There are initial states $\psi_0$ for which $P(\psi_0,L,\Phi,\epsilon,t) \sim C_\delta(\psi_0,L,\Phi,\epsilon)/t^{\delta}$, as $t \rightarrow \infty$, where $\delta = 1$ if $\cos(\Phi)=1$ and $\delta = 3$ if $\cos(\Phi)\not=1$. Thus, by submitting the loop to an external magnetic flux, one may induce a radical change in the asymptotic decay rate of $P(\psi_0,L,\Phi,\epsilon,t)$.  Interestingly, if $\cos(\Phi)=1$, then $C_1(\psi_0,L,\Phi,\epsilon)$ decreases with $\epsilon$ (\ie the particle escapes faster in the long run) while in the case $\cos(\Phi)\not=1$, the coefficient $C_3(\psi_0,L,\Phi,\epsilon)$ increases with $\epsilon$ (\ie the particle escapes slower in the long run). Assuming the particle to be initially in a bound state of the loop with $\Phi=0$, we  compute explicit relations and present some numerical results showing a global picture in time of $P(\psi_0,L,\Phi,\epsilon,t)$. Finally, by using the pseudo-spectral method, we consider the interacting case with soft-core Coulomb potentials.
\end{abstract}


\maketitle


\section{Introduction}

The study of open quantum systems has attracted many researchers in theoretical physics during the last decades and is still under intense investigation. 
One of the most basic problems in this field is to understand how a quantum particle initially confined inside an open cavity will escape (see \eg \cite{Exner-book} for a general discussion). 
 In the literature, such systems are commonly described in terms of the following quantities: the \emph{nonescape probability} $P(t)$, which is the probability that the particle will still be inside the cavity at some later time $t$ and the \emph{survival probability} $S(t)$, which is the probability to find the particle in the initial state at time $t$ \cite{CommentS}. Let us suppose that the configuration space is the one-dimensional half-line $[0,\infty)$ and that the cavity is the interval $[0,L]$. Then, at time $t=0$, the particle is supposed to be in a normalized state $\psi_0$ satisfying $\psi_0(x)=0$ for all $x \not \in [0,L]$. If $H$ denotes the Hamiltonian of the particle and $\psi(x,t)=(e^{-iHt/\hbar} \psi_0)(x)$ its state at time $t > 0$, where $\hbar$ is the reduced Planck constant, then the nonescape and survival probabilities are given by
\begin{eqnarray}
P(t) &=& \int_0^L \overline{\psi(x,t)} \psi(x,t) dx~,\\
S(t) &=& \left|\int_0^L \overline{\psi_0(x)} \psi(x,t) dx\right|^2~,
\end{eqnarray} 
where the bar "$\bar{\hspace{4mm}}$" denotes complex conjugation. It is easy to check that these quantities satisfy $S(0)=P(0)=1$ and $0 \leq S(t) \leq P(t) \leq 1$ for all times $t > 0$. Although both quantities are interesting, we shall discuss only $P(t)$ in this paper. Note, however, that by analysing the time decay of $P(t)$ one automatically gets an upper bound for $S(t)$.

The nonescape probability $P(t)$ has been discussed in free 1D quantum systems \cite{Miyamoto}, in free 2D quantum billiard systems \cite{Alt,Zozoulenko} and also in 3D quantum systems with finite range potentials \cite{AmreinControversy}. Interestingly, if the initial state $\psi_0$ has no bound state components, then it is always found that the long-time behaviour of $P(t)$ is a power law: $P(t)=C/t^\delta + \mathcal{O}(1/t^{\delta+1})$ as $t \rightarrow \infty$, where the constant $C$ and the exponent $\delta$ depend on the geometry of the system and on the initial state $\psi_0$ of the particle. Some readers may be surprised since it is well known that various spontaneous decays have been observed to follow an exponential law. This apparent discrepancy may actually be only a time scale problem. Indeed, although quantum mechanics clearly predicts in general some deviation from exponential decay, the following phases may well occur \cite{Ballentine,Wilkinson,Rothe}: for a short time the decay is parabolic $P(t) = 1 - c t^2$, then it is approximately exponential and finally becomes a power-law at very large times. 

\begin{figure}[htbp]
\begin{center}
\includegraphics[width=8cm]{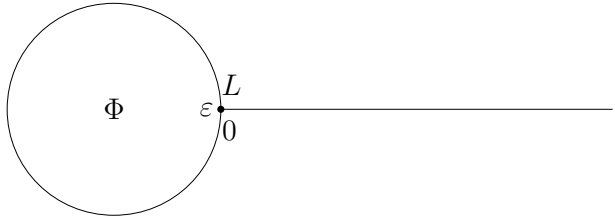}
\caption{A loop connected to a lead.}\label{Fig1}
\end{center}
\end{figure}

In the present paper, we consider the model introduced in \cite{Buttiker-Small} in which the cavity is a one-dimensional loop $[0,L]$ crossed by a constant magnetic flux $\Phi$ and connected, with coupling parameter $\epsilon$, to an infinite one-dimensional lead $[L,\infty)$ (the points $x=0$ and $x=L$ are identified), so that the particle may escape freely from the loop. See \Fig\ref{Fig1}. This model contains rich physics and is simple enough to be amenable by analytical means or numerical simulations, and thus has been the subject of several works \cite{Buttiker-Small,Buttiker-flux-sensitive,Akkermans,Xia,Jayannavar,Buttiker-char,Exner,Buttiker-suppression,Benjamin}. Nevertheless, the nonescape probability $P(\psi_0,L,\Phi,\epsilon,t)$ has never been discussed in this model and it is interesting to know precisely how it depends on the initial state $\psi_0$ of the particle and on the three physical parameters: the loop's length $L$, the external magnetic flux $\Phi$ and the coupling $\epsilon$.

In appropriate units, here is a summary of our main results.
\begin{itemize}
\item Let $\lambda > 0$, then one has the following scaling law:
\begin{equation}
P(\psi_0,\lambda L,\Phi,\epsilon,t)=P(\psi_0,L,\Phi,\epsilon,\frac{t}{\lambda^2})~.
\end{equation}
Thus, increasing for example the length of the loop by a factor $\lambda=2$ leads to a nonescape probability evolving $\lambda^2=4$ times slower.
\item The nonescape probability may be written as
\begin{equation}
P(\psi_0,L,\Phi,\epsilon,t) = P_\infty(\psi_0,\Phi) + \sum_{k=1}^\infty \frac{C_k(\psi_0,L,\Phi,\epsilon)}{t^k}~.
\end{equation}
The constant $P_\infty(\psi_0,\Phi)$ is given in \eref{Pinfinity}: it is independent of $L$ and $\epsilon$, and vanishes if $\psi_0$ has no bound state components or if $|\cos(\Phi)|\not=1$. In the infinite series, the leading term always has an odd power. The coefficients $C_1(\psi_0,L,\Phi,\epsilon)$ and $C_3(\psi_0,L,\Phi,\epsilon)$ are given in \eref{Equ for C1}--\eref{Equ for C3}: they depend on the initial state $\psi_0$ of the particle, but only the momentum $k=\Phi/L$ is involved. When $\Phi=0$, this coincides with the well-known fact that only the zero momentum of $\psi_0$ plays a role in the long-time behaviour of the nonescape probability \cite{Miyamoto}. If $\cos(\Phi)=1$, then there are initial states $\psi_0$ with no bound state components [\eg \eref{Initial State No Bound}] for which $C_1(\psi_0,L,\Phi,\epsilon)\not=0$, while the coefficients $C_1(\psi_0,L,\Phi,\epsilon)$ and $C_2(\psi_0,L,\Phi,\epsilon)$ always vanish if $\cos(\Phi)\not =1$. The important consequence is that there are initial states $\psi_0$ giving a power law decay: 
\begin{equation}
P(\psi_0,L,\Phi,\epsilon,t)=\frac{C_\delta(\psi_0,L,\Phi,\epsilon)}{t^\delta} + \mathcal{O}\left(\frac{1}{t^{\delta+1}}\right)~,
\end{equation}
where $\delta = 1$ if $\cos(\Phi)=1$ and $\delta = 3$ if $\cos(\Phi)\not=1$. In particular, we see that by submitting the loop to an external magnetic flux (thus going from $\Phi=0$ to $\Phi\not=0$), one may induce a radical change in the decay rate of $P(\psi_0,L,\Phi,\epsilon,t)$. If $\cos(\Phi)=1$, then $C_1(\psi_0,L,\Phi,\epsilon)$ decreases with $\epsilon$ (\ie the particle escapes faster in the long run) and if $\cos(\Phi)\not=1$, then $C_3(\psi_0,L,\Phi,\epsilon)$ increases with $\epsilon$ (\ie the particle escapes slower in the long run). Interestingly, similar features were obtained in a circular dielectric cavity containing classical waves \cite{Ryu}, the nature of the waves (TM or TE) and the refractive index of the cavity playing a similar role than $\Phi$ and $\epsilon$.
\item Assuming the particle to be initially in a bound state of the loop with $\Phi=0$ [see \eref{Bound states}], we computed explicitly $P_\infty(\psi_0,\Phi)$, $C_1(\psi_0,L,\Phi,\epsilon)$ and $C_3(\psi_0,L,\Phi,\epsilon)$ [see \eref{Pasymp}--\eref{Coeff C3 two}]. As one may expect, we found that higher energetic bound states decay faster. Interestingly, we observed that the coefficient $C_3(\psi_0,L,\Phi,\epsilon)$ oscillates, non-periodically, with $\Phi$; see \Fig\ref{Fig5}. We also present some numerical results showing a global picture in time of $P(\psi_0,L,\Phi,\epsilon,t)$; see \Figs\ref{Fig6}--\ref{Fig10}. They reveal in particular that if $\cos(\Phi)\not=1$, then by increasing the value of $\epsilon$, the particle escapes faster in the beginning but slower in the long run as stated previously.
\end{itemize}

This paper is organized as follows. In \Section~\ref{The Hamiltonian}, we model the situation depicted in \Fig\ref{Fig1} in the framework of standard quantum mechanics. In the paper \cite{Buttiker-Small}, the scattering processes occurring at the connecting point (between the loop and the lead) are described in terms of an energy-independent scattering matrix $S$. We shall show that this scattering matrix may be used to prescribe boundary conditions at the connecting point for which the Hamiltonian of the particle is self-adjoint.  In \Section~\ref{The propagator}, we solve the time-dependent Schr\"odinger equation and derive its associated propagator. We then analyse the long time behaviour of the propagator and consequently that of the nonescape probability. We discuss our general results in \Section~\ref{Discussion} and consider some particular initial states in \Section~\ref{Numerical Results}. In \Section~\ref{The Interacting case}, we present an interesting and highly non-trivial application of the pseudo-spectral method by considering the interacting case with soft-core Coulomb potentials. Finally, in \Section~\ref{Conclusion} we make some concluding remarks.

\section{The Hamiltonian}\label{The Hamiltonian}

In standard quantum mechanics, the situation depicted in \Fig\ref{Fig1} is modelled as follows: the physical states of the particle belong to the Hilbert space $\mathcal{H} = L^2([0,L]) \oplus L^2([L,\infty))$ and  
the Hamiltonian of the particle reads \cite{Buttiker-Small}
\begin{equation}\label{Hamiltonian}
H = \left[-i\frac{d}{dx} - \frac{\Phi}{L} \chi_{[0,L]}(Q)\right]^2~,
\end{equation}
where $\chi_{[0,L]}(Q)$ is the operator of multiplication by $\chi_{[0,L]}(x)$, where $\chi_{[0,L]}(x) = 1$ if $x \in [0,L]$ and  $\chi_{[0,L]}(x) = 0$ otherwise. Here we have set $\hbar=1$ for the reduced Planck constant, $e=1$ for the electric charge, $c=1$ for the speed of light and $m=1/2$ for the mass of the particle. 

Let $\psi$ be a state in $\mathcal{H}$, then it can be uniquely written as $\psi = \psi_{\scriptscriptstyle \rm LOOP} + \psi_{\scriptscriptstyle \rm LEAD}$, with $\psi_{\scriptscriptstyle \rm LOOP} \in L^2([0,L])$ and $\psi_{\scriptscriptstyle \rm LEAD} \in L^2([L,\infty))$. To simplify the notations, we shall use the following convention: $\psi(x)=\psi_{\scriptscriptstyle \rm LOOP}(x)$ if $x \in (0,L)$, $\psi(x)=\psi_{\scriptscriptstyle \rm LEAD}(x)$ if $x \in (L,\infty)$ and $\psi(0_+)=\psi_{\scriptscriptstyle \rm LOOP}(0)$, $\psi(L_-)=\psi_{\scriptscriptstyle \rm LOOP}(L)$, $\psi(L_+)=\psi_{\scriptscriptstyle \rm LEAD}(L)$.

\begin{figure}[htbp]
\centerline{\includegraphics[width=8cm]{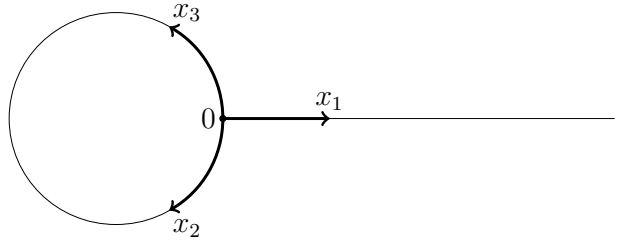}}
\caption{The local reference coordinates.}\label{Fig2}
\end{figure}

The Hamiltonian $H$ is an unbounded operator and thus cannot be defined on the entire Hilbert space $\mathcal{H}$. We thus have to find the domain $D(H) \subset \mathcal{H}$ corresponding to the situation depicted in \Fig\ref{Fig1} such that $H \psi \in \mathcal{H}$ for all $\psi \in D(H)$ and $\{H,D(H)\}$ is self-adjoint. The precise definition of a self-adjoint operator is given in Appendix~A and we refer to the paper \cite{Araujo} for some physical motivations. This section is not crucial to understand the remaining of this paper, so the uninterested reader may look at the scattering matrix \eref{Scattering matrix Markus} and then go directly to the solution \eref{Domain of H}. In the paper \cite{Buttiker-Small}, the scattering processes occurring at the connection point (between the loop and the lead) are described in terms of the following energy-independent (unitary) scattering matrix:
\begin{equation}\label{Scattering matrix Markus}
S = \matrice{ -(a+b) & \sqrt{\epsilon} & \sqrt{\epsilon} \\ \sqrt{\epsilon} & a & b \\ \sqrt{\epsilon} & b & a}~,
\end{equation}
where $a=\frac{1}{2} (\sqrt{1-2\epsilon}-1)$, $b=\frac{1}{2} (\sqrt{1-2\epsilon}+1)$ and $\epsilon \in (0,\frac{1}{2}]$. Here, $\epsilon = 0$ corresponds to the uncoupled situation (which is excluded) and $\epsilon = \frac{1}{2}$ to the maximally coupled one. Note that $a$ and $b$ never vanish. To implement such a scattering matrix, it is convenient to work in the local reference coordinates $\{x_1 \in [0,\infty) \mbox{ and } x_2,x_3 \in [0,L]\}$
associated to the three branches exiting from the connection point $x_1=x_2=x_3=0$. See \Fig\ref{Fig2}. Then, a general scattering matrix $S(k)$ at energy $k^2 > 0$, with $k > 0$, relates the incoming and outgoing amplitudes at the connection point through the following solutions of the stationary Schr\"odinger equation $H \varphi^\ell = k^2 \varphi^\ell$, with $k>0$ and $\ell = 1,2,3$:
\begin{eqnarray}
\varphi_1^\ell(x_1) &=& \delta_{1 \ell} e^{-ikx_1} + S_{1 \ell}(k) e^{ikx_1}~,\\
\varphi_2^\ell(x_2) &=& \left[\delta_{2 \ell} e^{-ikx_2} + S_{2 \ell}(k) e^{ikx_2}\right] e^{i\frac{\Phi}{L}x_2}~,\\
\varphi_3^\ell(x_3) &=& \left[\delta_{3 \ell} e^{-ikx_3} + S_{3 \ell}(k) e^{ikx_3}\right] e^{-i\frac{\Phi}{L}x_3}~,
\end{eqnarray}
where $\delta_{ij}$ is the Kronecker delta. Inspired from quantum networks studies \cite{Kuchment, Schrader-1,Schrader-2,Schrader-3}, we shall now derive the corresponding boundary conditions at the connection point insuring the self-adjointness of the Hamiltonian $H$.

Let $f$ and $g$ be two functions in $\mathcal{H}$ such that $Hf$ and $Hg$ are still in $\mathcal{H}$. Then, a simple calculation gives
\begin{equation}
\langle f | H g \rangle = \langle H f | g \rangle + \mbox{BT}~,
\end{equation}
where $\langle \cdot | \cdot \rangle$ denotes the usual scalar product in $L^2([0,\infty))$ and $\mbox{BT}$ contains the boundary terms appearing during the integrations by parts: 
\begin{equation}\label{BT local}
\mbox{BT} = \sum_{j=1}^3 \left[ \overline{f_j(0)} g_j'(0) - \overline{f_j'(0)} g_j(0) \right] + 2 i \frac{\Phi}{L} \left[ \overline{f_3(0)} g_3(0) - \overline{f_2(0)} g_2(0) \right]~,
\end{equation}
where $f_1(x_1)=f(L+x_1)$, $f_2(x_2)=f(x_2)$, $f_3(x_3)=f(L-x_3)$ and similarly for $g_1$, $g_2$ and $g_3$. The self-adjointness of $H$ requires that $\mbox{BT}$ vanishes for all $\mathbf{f}=(f_1,f_2,f_3)$ and $\mathbf{g}=(g_1,g_2,g_3)$ satisfying \emph{the same} boundary conditions at $x_1=x_2=x_3=0$ (see Appendix~A for a more precise statement). Setting $\mathbf{f}=\mathbf{g}$, one thus has the following \emph{necessary} condition:
\begin{equation}\label{Equation to be satisfied}
\langle \mathbf{f}(0)| \mathbf{f}'(0) \rangle_{\mathbb{C}^3} - \langle \mathbf{f}'(0)| \mathbf{f}(0) \rangle_{\mathbb{C}^3} +  2 i \frac{\Phi}{L} \left[|f_3(0)|^2 - |f_2(0)|^2 \right] = 0~,
\end{equation}
where $\langle \cdot | \cdot \rangle_{\mathbb{C}^3}$ denotes the canonical scalar product in $\mathbb{C}^3$. If $||\cdot||_{\mathbb{C}^3} = \sqrt{\langle \cdot | \cdot \rangle_{\mathbb{C}^3}}$, then  \eref{Equation to be satisfied} is satisfied if and only if $||\mathbf{f}(0) + i [\mathbf{f}'(0)-i\frac{\Phi}{L} \mathbf{h}(0)]||_{\mathbb{C}^3} = ||\mathbf{f}(0) - i [\mathbf{f}'(0)-i\frac{\Phi}{L} \mathbf{h}(0)]||_{\mathbb{C}^3}$, where $\mathbf{h}(0)=(0,f_2(0),-f_3(0))$. This is the case if and only if there is a unitary matrix $U$ such that
\begin{equation}
\mathbf{f}(0) + i \left[\mathbf{f}'(0)-i\frac{\Phi}{L} \mathbf{h}(0)\right] = U \left\{\mathbf{f}(0) - i \left[\mathbf{f}'(0)-i\frac{\Phi}{L} \mathbf{h}(0)\right]\right\}~,
\end{equation}
or
\begin{equation}\label{Boundary condition with U}
(\mathds{1}-U) \mathbf{f}(0) + i (\mathds{1}+U) \left[\mathbf{f}'(0)-i\frac{\Phi}{L} \mathbf{h}(0)\right] = 0~.
\end{equation}
We thus have the following \emph{necessary} condition for the self-adjointness of $\{H,D(H)\}$: there must be a unitary matrix $U$ such that the boundary conditions \eref{Boundary condition with U} are satisfied for all $f \in D(H)$. 

To specify the boundary conditions \eref{Boundary condition with U} in terms of the scattering matrix $S(k)$, one requires that the scattering states $\varphi^\ell$ (which are not in $\mathcal{H}$) satisfy the relation \eref{Boundary condition with U}. Setting $f = \varphi^\ell$, with $\ell =1,2$ or $3$, one easily checks that the relation \eref{Boundary condition with U} is verified with $U=[(k+1)S(k)-(k-1)\mathds{1}]\cdot[(k+1)\mathds{1}-(k-1)S(k)]^{-1}$, which means that the scattering matrix may be written as $S(k)=[(k-1)\mathds{1}+(k+1)U]\cdot[(k+1)\mathds{1}+(k-1)U]^{-1}$. Note that the knowledge of the scattering matrix $S(k)$ at any energy $E=k^2$ uniquely determines the boundary conditions \eref{Boundary condition with U} and that if $S$ does not depend on $k$, then one simply has $U=S$. Returning to our original reference frame, the scattering matrix \eref{Scattering matrix Markus} leads to the following boundary conditions:
\begin{eqnarray}
 \psi(0_+) - \psi(L_-) &=& 0 ~, \label{Condition 1}\\
 \sqrt{\epsilon} \psi(0_+) - b  \psi(L_+) &=& 0 ~, \label{Condition 2}\\
\sqrt{\epsilon} \psi'(L_+) +b \left[\psi'(0_+) - \psi'(L_-)\right] &=& 0 \label{Condition 3}~.
\end{eqnarray}
Note that these conditions do not depend on the magnetic flux $\Phi$, that they are contained in the model considered in \cite{Exner} and that the free boundary conditions $\psi(0_+) = \psi(L_-) = \psi(L_+)$ and $\psi'(L_-) = \psi'(0_+) + \psi'(L_+)$ correspond to the case $\epsilon = 4/9$. As explained previously, the boundary conditions \eref{Condition 1}-\eref{Condition 3} are \emph{necessary} for $H$ to be self-adjoint. In Appendix~A, we show that these boundary conditions are also sufficient for $H$ to be self-adjoint. Therefore,  we deduce that the relations \eref{Condition 1}-\eref{Condition 3} are in fact the \emph{unique} boundary conditions associated to the scattering matrix \eref{Scattering matrix Markus} for which the Hamiltonian \eref{Hamiltonian} is self-adjoint. More precisely, the Hamiltonian $H$ associated to $S$ is self-adjoint if and only if 
\begin{equation}\label{Domain of H}
D(H) \hspace{-1mm} = \hspace{-1mm} \{ \psi \in \mathcal{H} \ | \ H \psi \in \mathcal{H} \mbox{ and }  \eref{Condition 1}-\eref{Condition 3} \mbox{ are satisfied} \}
\end{equation}
 
\section{The propagator}\label{The propagator}

Let us recall that the model we are describing contains three parameters: the loop's length $L > 0$, the magnetic flux $\Phi \in \real$ and the coupling parameter $\epsilon \in (0,\frac{1}{2}]$. In what follows, we shall suppose that their values are kept constant and that at time $t=0$ the initial state $\psi_0 \in D(H)$ of the particle is localized in the loop, \ie it satisfies $\psi_0(x) = 0$, for all $x \not \in [0,L]$, and is normalized: $\int_0^L |\psi_0(x)|^2 dx = 1$. Note that the property $\psi_0(L_+) = \psi'_0(L_+) = 0$ together with the boundary conditions \eref{Condition 1}--\eref{Condition 3} imply $\psi_0(0_+) = \psi_0(L_-) = 0$ and $\psi'_0(0_+) = \psi'_0(L_-)$. Then, 
the nonescape probability is given by ($t > 0$)
\begin{equation}\label{nonescape prob}
P(\psi_0,L,\Phi,\epsilon,t) = \int_0^L  |\psi(x,t)|^2 dx~,
\end{equation}
where the wave function obeys the Schr\"odinger equation
\begin{equation}\label{Schrodinger equ}
i  \frac{\partial}{\partial t} \psi(x,t) = H  \psi(x,t)~, \hspace{3mm} \mbox{ with } \hspace{3mm} \psi(x,0) = \psi_0(x)~.
\end{equation}
We see that we need to solve the Schr\"odinger equation \eref{Schrodinger equ} inside the loop ($x \in [0,L]$) in order to compute the nonescape probability \eref{nonescape prob}. To determine $\psi(x,t)$ we shall write the initial state $\psi_0(x)$ as a superposition of generalized eigenstates of $H$, satisfying \eref{Condition 1}--\eref{Condition 3}, over the spectrum of $H$, which is $\sigma(H)= [0,\infty)$ (see Appendix~A).

Let $k > 0$, then the general solution of the time-independent Sch\"odinger equation $H \varphi_k = k^2 \varphi_k$ is
 \begin{eqnarray}
\varphi_{k}(x) &=& \left[A(k) e^{i k x} +  B(k) e^{-i k x}\right] e^{i \frac{\Phi}{L}x}  \chi_{[0,L)}(x)\nonumber\\
 & & + \left[C(k) e^{i k (x-L)} +  D(k) e^{-ik (x-L)}\right] \chi_{[L,\infty)}(x)~,\label{Stationary solution}
\end{eqnarray}
where $A(k)$, $B(k)$, $C(k)$ and $D(k)$ are complex numbers. We shall choose these constants such that the boundary conditions \eref{Condition 1}--\eref{Condition 3} are satisfied and 
such that the generalized eigenstates $\{\varphi_{k}(x)\}$ are $\delta$-normalized (as we shall see $\varphi_{-k}(x)=-\varphi_{k}(x)$, so only the $\varphi_{k}(x)$ with positive $k$ need to be considered): 
\begin{equation}\label{Completness}
\langle \varphi_{k} | \varphi_{p} \rangle = \int_0^\infty \overline{\varphi_{k}(x)} \varphi_{p}(x) dx   = \delta(k - p)~, \ \mbox{ for all } \ k, p > 0~.
\end{equation} 
Solving \eref{Condition 1}--\eref{Condition 3} and \eref{Completness}, one finds \cite{Comment1}
\begin{eqnarray}
A(k) &=& \frac{e^{-i (k L - \Phi)} - 1}{\mbox{DEN}(k)}~,\label{Coeff A}\\
B(k) &=& - \frac{e^{i (k L + \Phi)} - 1}{\mbox{DEN}(k)}~,\\
C(k) &=& \frac{1}{2 b \sqrt{\epsilon}} \ \frac{(2b-\epsilon) (1 + e^{2i\Phi}) - 2(b-\epsilon) e^{-i (k L - \Phi)} - 2b e^{i (k L + \Phi)}}{\mbox{DEN}(k)}~,\\
D(k) &=& -\frac{1}{2 b \sqrt{\epsilon}} \ \frac{(2b-\epsilon) (1 + e^{2i\Phi}) - 2(b-\epsilon) e^{i (k L + \Phi)} - 2b e^{-i (k L - \Phi)}}{\mbox{DEN}(k)}~,\label{Coeff D}
\end{eqnarray}
where
\begin{equation}\label{DEN}
\mbox{DEN}(k) = \left\{\frac{8 \pi}{\epsilon}  \left\{a^2 \sin^2(kL) + b^2 \left[\cos(kL) - \cos(\Phi)\right]^2 \right\} \right\}^{1/2}~.
\end{equation}
In Appendix~B, we show that the coefficients $A$, $B$, $C$ and $D$ are well defined in $\real$. In addition, if $\cos(\Phi)=1$ or $\cos(\Phi)=-1$, there are the bound states $\tilde{\varphi}^+_n, \tilde{\varphi}^-_n \in D(H)$ of $H$ ($C=D=0$ and $A=-B$): 
\begin{eqnarray}
\hspace{-10mm}\cos(\Phi)=1 &:& \tilde{\varphi}^+_n(x) = \sqrt{\frac{2}{L}} \sin\left(k_{2n} x\right) e^{i \frac{\Phi}{L} x} \ \chi_{[0,L)}(x)~, \hspace{2mm} n = 1, 2, \dots~,\label{Bound 1}\\
\hspace{-10mm}\cos(\Phi)=-1 &:& \tilde{\varphi}^-_n(x) = \sqrt{\frac{2}{L}} \sin\left(k_{2n+1} x\right) e^{i \frac{\Phi}{L} x} \ \chi_{[0,L)}(x)~, \hspace{2mm} n = 0, 1, \dots\label{Bound 2}
\end{eqnarray}
where $k_n = \frac{n \pi}{L}$. Note that the eigenvalues $E_n = \left(\frac{n \pi}{L}\right)^2$, with $n=1,2,\dots$, associated to the bound states \eref{Bound 1}--\eref{Bound 2} are as expected inside the continuous spectrum $[0,\infty)$ of $H$. If $\cos(\Phi)=1$, then one has for all $m, n = 1, 2, \dots$ and every $k >0$ ($k \in \{\frac{2n \pi}{L}\}_{n=1}^{\infty}$ included):
\begin{equation}\label{Relation bound-scattering}
\langle \tilde{\varphi}^+_{m} | \tilde{\varphi}^+_{n} \rangle = \delta_{m  n} \hspace{6mm} \mbox{and} \hspace{6mm} \langle \varphi_{k} | \tilde{\varphi}^+_{n} \rangle = 0~.
\end{equation}
Similarly, if $\cos(\Phi)=-1$, then for all $m, n = 0, 1, \dots$ and $k >0$ ($k \in \{\frac{(2n+1) \pi}{L}\}_{n=0}^{\infty}$ included):
\begin{equation}\label{Relation bound-scattering2}
\langle \tilde{\varphi}^-_{m} | \tilde{\varphi}^-_{n} \rangle = \delta_{m  n} \hspace{6mm} \mbox{and} \hspace{6mm} \langle\varphi_{k} | \tilde{\varphi}^-_{n} \rangle = 0~.
\end{equation}
Since the Hamiltonian $H$ is self-adjoint, it follows from the Spectral Theorem that the family of states $\{\varphi_k, \delta_{\cos(\Phi),1} \cdot \tilde{\varphi}^+_n, \delta_{\cos(\Phi),-1} \cdot \tilde{\varphi}^-_n\}$ generates the whole Hilbert space $\mathcal{H}$. To avoid technical complications, we shall assume that the initial state $\psi_0$ can be written \emph{punctually} in terms of these states [this assumption may be checked explicitly in the cases $(\epsilon = 1/2, \Phi = 0)$ and $(\epsilon = 1/2, \Phi = \pi/2)$].\\
{\bf Assumption (completeness):} Let $\psi_0 \in D(H)$ be an initial state localized in the loop. Then, we assume that for (almost) every $x \in [0,\infty)$:
\begin{equation}\label{Equ completeness}
\psi_0(x) = \int_0^\infty \langle \varphi_{k} | \psi_0 \rangle \varphi_{k}(x) dk + \delta_{\cos(\Phi),1} \cdot \sum_{n=1}^{\infty}\langle \tilde{\varphi}^+_{n} | \psi_0 \rangle \tilde{\varphi}^+_{n}(x) + \delta_{\cos(\Phi),-1} \cdot \sum_{n=0}^{\infty} \langle \tilde{\varphi}^-_{n} | \psi_0 \rangle \tilde{\varphi}^-_{n}(x)~.
\end{equation}

Under this assumption and using the orthogonality relations \eref{Completness} and \eref{Relation bound-scattering}--\eref{Relation bound-scattering2}, one may write the solution of the Schr\"odinger equation \eref{Schrodinger equ} as
\begin{equation}\label{The time-dependent wave function}
\psi(x,t) = (e^{-iHt} \psi_0)(x) = \int_0^L K(x,y,t) \psi_0(y) dy~,
\end{equation}
where the propagator is
\begin{eqnarray}\label{propagator}
K(x,y,t) &=& \int_0^\infty e^{-ik^2t} \varphi_k(x) \overline{\varphi_k(y)} dk\nonumber\\
&& + \delta_{\cos(\Phi), 1} \cdot \sum_{n=1}^{\infty} e^{-i k^2_{2n} t} \tilde{\varphi}^+_n(x) \overline{\tilde{\varphi}^+_n(y)} \nonumber\\
&& + \delta_{\cos(\Phi), -1} \cdot \sum_{n=0}^{\infty} e^{-i k^2_{2n+1} t} \tilde{\varphi}^-_n(x) \overline{\tilde{\varphi}^-_n(y)} ~.
\end{eqnarray}
Looking back at \eref{nonescape prob} and \eref{The time-dependent wave function} one sees that it is sufficient to know the propagator $K(x,y,t)$ for $x,y \in [0,L)$. In this case, one may write
\begin{equation}\label{Formula for the propagator}
K(x,y,t) = K_1(x,y,t) + \delta_{\cos(\Phi),1} \cdot K_2(x,y,t) +  \delta_{\cos(\Phi),-1} \cdot K_3(x,y,t) ~,
\end{equation}
where
\begin{eqnarray}
\hspace{-9mm}K_1(x,y,t) &=& e^{i\frac{\Phi}{L}(x-y)} \int_{-\infty}^\infty \left\{f_1(k) e^{-i [k^2 t - k (x-y)]} + f_2(k) e^{-i [k^2 t - k (x+y)]}\right\} \ dk~, \label{propagator details1}\\
\hspace{-9mm}K_2(x,y,t) &=&  \frac{e^{i \frac{\Phi}{L}(x-y)}}{2L}  \sum_{n=-\infty}^{\infty} \left\{e^{-i \left[k^2_{2n} t - k_{2n}^{} (x-y)\right]} -  e^{-i \left[k_{2n}^2 t - k_{2n}^{} (x+y)\right]}\right\}~,\label{propagator details2}\\
\hspace{-9mm}K_3(x,y,t) &=&  \frac{e^{i \frac{\Phi}{L}(x-y)}}{2L}  \sum_{n=-\infty}^{\infty} \left\{e^{-i \left[k_{2n+1}^2 t - k_{2n+1}^{} (x-y)\right]} -  e^{-i \left[k_{2n+1}^2 t - k_{2n+1}^{} (x+y)\right]}\right\}~,\label{propagator details3}
\end{eqnarray}
with
\begin{eqnarray}
k_n &=& \frac{n \pi}{L}~, \hspace{5mm} n \in \mathbb{Z}~,\\
f_1(k) &=&  \frac{\epsilon \left[1-\cos(kL-\Phi) \right]}{4 \pi  \left\{a^2 \sin^2(kL) + b^2 \left[\cos(kL) - \cos(\Phi)\right]^2 \right\}}~,\label{Fonction f1}\\
f_2(k) &=& -\frac{\epsilon \left[1+e^{-2ikL} - 2 e^{-ikL} \cos(\Phi) \right]}{8 \pi  \left\{a^2 \sin^2(kL) + b^2 \left[\cos(kL) - \cos(\Phi)\right]^2 \right\}}~.\label{Fonction f2}
\end{eqnarray}
Clearly, the propagator $K_1(x,y,t)$ describes the decaying aspect of the initial state $\psi_0$, while the two other propagators, $K_2(x,y,t)$ and $K_3(x,y,t)$, describe the stationary aspect of $\psi_0$, \ie the part of $\psi_0$ which remains in the loop for all times \cite{Bound}.

In the case ($\epsilon = 1/2$, $\Phi = \pi/2$), one has $\mbox{DEN}(k) = \sqrt{4 \pi}$ and one may evaluate the propagator exactly by splitting the different terms. We find
\begin{equation}\label{Exact propagator 1}
K(x,y,t)^{\hspace{-15mm}\begin{array}{l}\vspace{-4.5mm} \scriptstyle(\epsilon = \frac{1}{2}, \Phi = \frac{\pi}{2})\\ \ \end{array}} \hspace{-3mm} = \frac{e^{i\frac{\pi}{2L}(x-y)}}{\sqrt{16 \pi i t}}  \left[2e^{i\frac{(x-y)^2}{4t}} - e^{i\frac{(x+y)^2}{4t}} -  e^{i\frac{(x+y-2L)^2}{4t}} +i e^{i\frac{(x-y+L)^2}{4t}} - i e^{i\frac{(x-y-L)^2}{4t}}\right]~.
\end{equation}
Also, in the case ($\epsilon = 1/2$, $\Phi = 0$), one has $f_1(k)=1/(4 \pi)$ and $f_2(k)=e^{-ikL}/(4 \pi)$, so that
\begin{equation}\label{Exact propagator 2}
K(x,y,t)^{\hspace{-15mm}\begin{array}{l}\vspace{-4.5mm} \scriptstyle(\epsilon = \frac{1}{2}, \Phi = 0)\\ \ \end{array}} \hspace{-2mm} = \frac{1}{\sqrt{16 \pi i t}}  \left[e^{i\frac{(x-y)^2}{4t}} + e^{i\frac{(x+y-L)^2}{4t}}\right] + K_2(x,y,t)~.
\end{equation}
In general, we are not able to integrate explicitly $K_1(x,y,t)$, so we shall only discuss its long time behaviour (some comments concerning the short-time behaviour of $P(\psi_0,L,\Phi,\epsilon,t)$ can be found in \Section~\ref{Conclusion}). For this, we need to analyse the asymptotic behaviour of an integral of the following form:
\begin{equation}\label{Equ for I}
I(t) = \int_{-\infty}^{\infty} f(k) e^{-i (k^2 t - k z)} dk~,
\end{equation}
where $f$ is either $f_1$ or $f_2$ and $z = (x \pm y) \in \real$. Looking at the expressions \eref{Fonction f1}--\eref{Fonction f2}, \eref{f1 cos1}--\eref{f2 cos1} and \eref{f1 cosm1}--\eref{f2 cosm1}, it is clear that, for any $k \in \mathbb{R}$, the function $f(k)$ may be written as a Maclaurin series:
\begin{equation}\label{Expansion of f}
f(k) = \sum_{n=0}^\infty \frac{f^{(n)}(0)}{n!} k^n~.
\end{equation}
Substituting this expression in \eref{Equ for I} and exchanging formally the order of integration and summation we obtain
\begin{equation}
I(t) = \sum_{n=0}^\infty \frac{f^{(n)}(0)}{n!} \int_{-\infty}^{\infty} k^n e^{-i (k^2 t - k z)} dk~.
\end{equation}
We next use the following identity (see Appendix~C):
\begin{equation}\label{Identity Gaussian Integral}
 \int_{-\infty}^{\infty} k^n e^{- a k^2 + b k} dk = \sqrt{\frac{\pi}{a}} \ e^{\frac{b^2}{4a}} \ \sum_{\ell = 0}^{\lfloor \frac{n}{2} \rfloor} \frac{n!}{\ell! \ (n-2\ell)!} \ \frac{(2b)^{n-2\ell}}{(4a)^{n-\ell}}~,
\end{equation}
where $\lfloor \frac{n}{2} \rfloor = n/2$ if $n$ is even and $(n-1)/2$ if $n$ is odd, and $a,b \in \mathbb{C}\setminus \{0\}$, with $\mbox{Re}(a)>0$. 
The cases $b=0$ and $\mbox{Re}(a)=0$ are obtained as the limiting expressions of the r.h.s.~of \eref{Identity Gaussian Integral} as $b \rightarrow 0$ and $\mbox{Re}(a) \rightarrow 0^+$, respectively. Thus, writing 
\begin{equation}
e^{i \frac{z^2}{4t}} = \sum_{m=0}^\infty \frac{1}{m!} \ \frac{i^m z^{2m}}{(4t)^m}
\end{equation}
we obtain
\begin{equation}\label{Integral for I2}
I(t) = \sum_{n=0}^\infty  \sum_{m=0}^\infty \sum_{\ell=0}^{\lfloor \frac{n}{2} \rfloor}  \frac{\sqrt{\pi} \ i^{m-\ell-1/2}}{2^{2m+n} \ m! \ \ell! \ (n-2\ell)!} \ f^{(n)}(0) \ z^{2(m-\ell)+n} \ \frac{1}{t^{[2(m+n-\ell)+1]/2}}~.
\end{equation}
Using the relation \eref{Integral for I2}, one can write 
\begin{eqnarray}\label{Sum for kernel}
K_1(x,y,t) &=&  e^{i\frac{\Phi}{L}(x-y)} \sum_{n=0}^\infty  \sum_{m=0}^\infty \sum_{\ell=0}^{\lfloor \frac{n}{2} \rfloor} \frac{\sqrt{\pi} \ i^{m-\ell-1/2}}{2^{2m+n} \ m! \ \ell! \ (n-2\ell)!} \ \frac{1}{t^{[2(m+n-\ell)+1]/2}}\nonumber \\
&& \cdot \left[f^{(n)}_1(0) \ (x-y)^{2(m-\ell)+n} + f^{(n)}_2(0) \ (x+y)^{2(m-\ell)+n}\right] ~,
\end{eqnarray} 
where the functions $f_1$ and $f_2$ are given by \eref{Fonction f1}--\eref{Fonction f2}. The first two leading terms are
\begin{eqnarray}
K_1(x,y,t) &=&  e^{i\frac{\Phi}{L}(x-y)} \bigg\{\sqrt{\frac{\pi}{i}} [f_1(0) + f_2(0)] \frac{1}{t^{1/2}} +  \big\{\frac{\sqrt{\pi \ i}}{4} [f_1(0) (x-y)^2 + f_2(0) (x+y)^2] \nonumber\\
&& \hspace{-27mm}+ \sqrt{\frac{\pi}{4 \ i}} \ [f_1'(0) (x-y) + f_2'(0) (x+y)]+  \frac{\sqrt{\pi}}{4 \ i^{3/2}} \ [f_1''(0) + f_2''(0)] \big\} \frac{1}{t^{3/2}} \bigg\} + \mathcal{O}\left(\frac{1}{t^{5/2}}\right)~.
\end{eqnarray}
A careful analysis of the expressions \eref{Fonction f1}--\eref{Fonction f2} shows that one has to deal with two cases (see also Appendix~B): $\cos(\Phi) = 1$ and $\cos(\Phi) \not= 1$. In particular, one has
\begin{eqnarray}
f_1(0) = f_2(0) &=& \frac{\epsilon}{8 \pi a^2} \hspace{4mm} \mbox{if} \hspace{4mm} \cos(\Phi) = 1~,\\
f_1(0) = - f_2(0) &=& \frac{\epsilon}{4 \pi b^2 [1-\cos(\Phi)]} \hspace{4mm} \mbox{if} \hspace{4mm} \cos(\Phi) \not = 1~.
\end{eqnarray}
As we shall see, the fact that $f_1(0) = f_2(0)$ in the first case, while $f_1(0) = - f_2(0)$ in the second, will lead to a drastic change in the decay properties of the nonescape probability. 

If $\cos(\Phi) = 1$, then 
\begin{eqnarray}
K(x,y,t)  &=& K_2(x,y,t) + \frac{\epsilon e^{i\frac{\Phi}{L}(x-y)}}{\sqrt{16 \pi i} \ a^2} \ \Bigg\{\frac{1}{t^{1/2}} +\label{Expansion Kernel General 2}\\
 && \hspace{18mm} + \frac{i}{4} \left[x^2 + y^2 - L(x+y) + \frac{L^2}{2} \left(\frac{b}{a}\right)^2\right]  \ \frac{1}{t^{3/2}}\Bigg\} + \mathcal{O}\left(\frac{1}{t^{5/2}}\right)~.\nonumber
\end{eqnarray}
If $\cos(\Phi) \not = 1$, then
\begin{eqnarray}
&&\hspace{-7mm}K(x,y,t) = \delta_{\cos(\Phi), -1} \ K_3(x,y,t) -\frac{i \epsilon \ e^{i\frac{\Phi}{L}(x-y)}}{8 \sqrt{\pi i} \ b^2} \cdot \label{Expansion Kernel General 1}\\
&&  \frac{\left\{L-[1-\cos(\Phi)+i\sin(\Phi)]x \right\}  \ \left\{L-[1-\cos(\Phi)-i\sin(\Phi)]y \right\} }{[1-\cos(\Phi)]^2} \ \frac{1}{t^{3/2}}+ \mathcal{O}\left(\frac{1}{t^{5/2}}\right)~.\nonumber
\end{eqnarray}
As one may easily check, the expression \eref{Expansion Kernel General 2} with ($\epsilon=1/2$,$\Phi=0$) and the expression \eref{Expansion Kernel General 1} with ($\epsilon=1/2$,$\Phi=\pi/2$)
coincide with the leading terms of \eref{Exact propagator 2} and \eref{Exact propagator 1}, respectively.

\section{General Results}\label{Discussion}

From the relations \eref{Expansion Kernel General 2}--\eref{Expansion Kernel General 1}, it follows that 
\begin{equation}
\lim_{t \rightarrow \infty} P(\psi_0,L,\Phi,\epsilon,t) = \left\{\begin{array}{ll} P_{\infty}(\psi_0,L,\Phi) & \mbox{if } |\cos(\Phi)| = 1 \\  0 & \mbox{otherwise} \end{array}\right.~,
\end{equation}
where
\begin{equation}\label{Pinfinity}
P_\infty(\psi_0,L,\Phi) = \int_0^L dx \left|\int_0^L dy \left[\delta_{\cos(\Phi),1} \ K_2(x,y,0) + \delta_{\cos(\Phi),-1} \ K_3(x,y,0)\right] \psi_0(y)\right|^2~.
\end{equation}
If $\cos(\Phi)=1$ [or $\cos(\Phi)=-1$], then the initial state $\psi_0$ may be written as a sum of two terms [see \eref{Equ completeness}], one associated to the scattering states $\varphi_k$ and the other to the bound states $\tilde{\varphi}^+_n$ [or $\tilde{\varphi}^-_n$]. The part associated to the bound states will lead to a non-zero value of $P_{\infty}(\psi_0,L,\Phi)$. Note that $P_{\infty}(\psi_0,L,\Phi)$ does not depend on $\epsilon$. From the expressions \eref{Expansion Kernel General 2}--\eref{Expansion Kernel General 1}, one sees that the large-time behaviour of $P(\psi_0,L,\Phi,\epsilon,t)$ depends drastically on the value of $\cos(\Phi)$. Nevertheless, setting $P_{\infty}(\psi_0,L,\Phi)=0$ whenever $|\cos(\Phi)| \not= 1$, one may write
\begin{equation}\label{Expression for P in series}
P(\psi_0,L,\Phi,\epsilon,t)  = P_{\infty}(\psi_0,L,\Phi) + \sum_{k=1}^\infty \frac{C_k(\psi_0,L,\Phi,\epsilon)}{t^{k}}~.
\end{equation}
To obtain explicit expressions for the coefficients $C_k(\psi_0,L,\Phi,\epsilon)$, it is convenient to write the expression \eref{Sum for kernel} in the following form:
\begin{equation}
K_1(x,y,t) = \sum_{\ell=0}^\infty \frac{G_{\ell}(x,y)}{t^{(2\ell+1)/2}}~.
\end{equation}
Then, the decaying part of the wave function reads
\begin{equation}
\psi_{\rm decay}(x,t) \equiv \int_0^L K_1(x,y,t) \psi_0(y)  dy = \sum_{\ell=0}^\infty \frac{Q_\ell(x)}{t^{(2\ell+1)/2}}~,
\end{equation}
where
\begin{equation}
Q_\ell(x) =  \int_0^L G_{\ell}(x,y) \psi_0(y) dy~.
\end{equation}
Hence, 
\begin{equation}
P(\psi_0,L,\Phi,\epsilon,t) - P_{\infty}(\psi_0,L,\Phi) = \int_0^L |\psi_{\rm decay}(x,t) |^2 dx = \sum_{k=1}^\infty \frac{C_k(\psi_0,L,\Phi,\epsilon)}{t^{k}}~,
\end{equation}
where
\begin{equation}
C_k(\psi_0,L,\Phi,\epsilon) =  \sum_{\scriptsize\begin{array}{c}
    \ell,m = 0\\ \ell+m+1=k \end{array}}^{\infty} \hspace{-3mm} \langle Q_m | Q_\ell \rangle~.
\end{equation}
Note that if $C_1(\psi_0,L,\Phi,\epsilon)=0$, then $Q_0 \equiv 0$ and thus $C_2(\psi_0,L,\Phi,\epsilon)=0$. More generally, if $C_1(\psi_0,L,\Phi,\epsilon)=C_2(\psi_0,L,\Phi,\epsilon)=\dots=C_{2\ell+1}(\psi_0,L,\Phi,\epsilon)=0$, then $C_{2\ell+2}(\psi_0,L,\Phi,\epsilon)=0$. Thus, at large times the nonescape probability $P(\psi_0,L,\Phi,\epsilon,t) \sim C_\delta(\psi_0,L,\Phi,\epsilon)/t^{\delta}$, with $\delta$ an \emph{odd} number. Let us use the expressions \eref{Expansion Kernel General 2}--\eref{Expansion Kernel General 1} to write the leading coefficients $C_k(\psi_0,L,\Phi,\epsilon)$. It turns out that they can be nicely written in terms of the derivatives of the Fourier transform of the initial state:
\begin{equation}
\tilde{\psi}^{(n)}_0(k) = \frac{(-i)^n}{\sqrt{2 \pi}} \int_{0}^L e^{-iky} y^n \psi_0(y) dy~.
\end{equation}
There are two cases:\\ \\
(1) If $\cos(\Phi) = 1$, then
\begin{equation}
C_1(\psi_0,L,\Phi,\epsilon) = \frac{\epsilon^2 L}{8 \ a^4} \left|\tilde{\psi}_0\left(\frac{\Phi}{L}\right)\right|^2~.\label{Equ for C1}
\end{equation}
If $C_1(\psi_0,L,\Phi,\epsilon)=0$, then $C_2(\psi_0,L,\Phi,\epsilon)=0$ and the next leading term is given by
\begin{equation}\label{Equ C3 when C1 is zero}
C_3(\psi_0,L,\Phi,\epsilon) = \frac{\epsilon^2 L}{128 \ a^4}  \left|L \tilde{\psi}'_0\left(\frac{\Phi}{L}\right) - i \tilde{\psi}''_0\left(\frac{\Phi}{L}\right)\right|^2~.
\end{equation}
(2) If $\cos(\Phi) \not= 1$, then $C_1(\psi_0,L,\Phi,\epsilon)=C_2(\psi_0,L,\Phi,\epsilon)=0$ and
\begin{eqnarray}
C_3(\psi_0,L,\Phi,\epsilon) &=& \frac{\epsilon^2 L^3 [2+\cos(\Phi)]}{96 \ b^4 [1-\cos(\Phi)]^4} \cdot \label{Equ for C3}\\
&&   \left|L  \tilde{\psi}_0\left(\frac{\Phi}{L}\right) - \{\sin(\Phi)+i[1-\cos(\Phi)]\} \tilde{\psi}'_0\left(\frac{\Phi}{L}\right)\right|^2\nonumber~.
\end{eqnarray}

{\bf $\boldsymbol{\psi_0}$-dependence:} We see that only the momentum $k=\Phi/L$ plays a role in the decaying properties of $P(\psi_0,L,\Phi,\epsilon,t)$. When $\Phi=0$, this coincides with the well-known fact that only the zero momentum is involved \cite{Miyamoto}. Note that $C_1(\psi_0,L,\Phi,\epsilon) \not = 0$ only if $\cos(\Phi) = 1$ and $\tilde{\psi}_0(\Phi/L) \not = 0$. As an illustration, let us discuss the case $\Phi = 0$. In this case, the bound states \eref{Bound 1} form a basis in $L^2([0,L])$ of all odd functions with respect to $L/2$ [\ie $\psi(x) = -\psi(L-x)$ for all $x \in [0,L]$]. Thus, if $\Phi = 0$ and the initial state $\psi_0 \in D(H)$ is an odd function with respect to $L/2$, then $P(\psi_0,L,\Phi,\epsilon,t)=P_{\infty}(\psi_0,L,\Phi)=1$ for all times. On the contrary, if $\Phi = 0$ and $\psi_0 \in D(H)$ is an even function with respect to $L/2$ [\ie $\psi_0(x) = \psi_0(L-x)$ for all $x \in [0,L]$], then $P_{\infty}(\psi_0,L,\Phi)=0$ and $P(\psi_0,L,\Phi,\epsilon,t) = C_1(\psi_0,L,\Phi,\epsilon)/t + \mathcal{O}(1/t^2)$ at large times, with $C_1(\psi_0,L,\Phi,\epsilon) \not = 0$ since $\int_0^L \psi_0(y) dy \not= 0$. More generally, if $\Phi = 0$ and $\psi_0 = \psi^{\rm odd}_0 + \psi^{\rm even}_0$, then $P(\psi_0,L,\Phi,\epsilon,t) = P_\infty(\psi_0,L,\Phi) + C_1(\psi_0,L,\Phi,\epsilon)/t + \mathcal{O}(1/t^2)$, 
where $P_\infty(\psi_0,L,\Phi) = \int_0^L dx |\int_0^L dy K_2(x,y,0) \psi^{\rm odd}_0(y)|^2$ and  $C_1(\psi_0,L,\Phi,\epsilon) = \epsilon^2 L /(8 a^4) \ |\tilde{\psi}^{\rm even}_0(0)|^2 $.

{\bf  $\boldsymbol{L}$-dependence:} Let $\psi_{0}(x,\lambda)$ and $K(x,y,t,\lambda)$ be the initial state and the propagator corresponding to a loop of length $\lambda L$, with $\lambda > 0$. Then, from the relations \eref{Bound states} and \eref{Formula for the propagator}, one deduces that
\begin{eqnarray}
\sqrt{\lambda} \ \psi_{0}(\lambda x,\lambda) &=& \psi_{0}(x,1)~,\\
\lambda \ K(\lambda x,\lambda y,\lambda^2 t,\lambda) &=& K(x,y,t,1)~.
\end{eqnarray}
From these scaling laws, one finds that 
\begin{equation}
P(\psi_0,\lambda L,\Phi,\epsilon,t) = P\left(\psi_0,L,\Phi,\epsilon,\frac{t}{\lambda^2}\right)~.
\end{equation}
From the relation \eref{Expression for P in series}, one easily deduces that $P_\infty$ is independent of $L$, so from now on we shall write $P_{\infty}(\psi_0,\Phi)$, and that the coefficient $C_k$ is an homogeneous function of degree $2k$, \ie $C_k(\psi_0,\lambda L,\Phi,\epsilon) = \lambda^{2k} \ C_k(\psi_0,L,\Phi,\epsilon)$ for all $\lambda > 0$, and thus $C_k(\psi_0,L,\Phi,\epsilon) \propto L^{2k}$.

{\bf  $\boldsymbol{\Phi}$-dependence:} Looking at the relations \eref{Equ for C1}--\eref{Equ for C3}, one sees that in general it is difficult to predict the $\Phi$-dependence of $P(\psi_0,L,\Phi,\epsilon,t)$ 
and we shall discuss this matter in detail in the next section. Nevertheless, some general symmetries can be found. Indeed, the propagators $K_j(x,y,t)$, with $j = 1, 2, 3$, become $e^{i \frac{2\pi}{L} (x-y)} K_j(x,y,t)$ under the transformation $\Phi \mapsto \Phi + 2 \pi$. This implies that $P(\psi_0,L,\Phi,\epsilon,t)$ is invariant under this transformation on the proviso that one replaces the initial state $\psi_{0}(y)$ by $e^{- i \frac{2\pi}{L} y} \psi_{0}(y)$. Furthermore, a simple calculation shows that $P(\psi_0,L,\Phi,\epsilon,t)$ is invariant under the transformation $\Phi \mapsto -\Phi$ if the initial state satisfies $\psi_0(x) = -\psi_0(L-x)$ [as in \eref{Bound states}] or $\psi_0(x) = \psi_0(L-x)$, for all $x \in [0,L]$. 

\begin{figure}[htbp]
\centerline{\includegraphics[width=8cm]{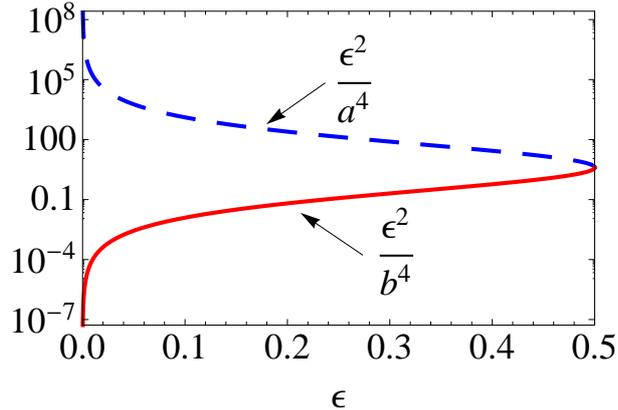}}
\caption{Log-lin plot of $\epsilon^2 / a^4$ (dashed blue) and $\epsilon^2 / b^4$ (red) for $\epsilon \in [0,0.5]$.}\label{Fig3}
\end{figure}

{\bf $\boldsymbol{\epsilon}$-dependence:} At large times, one sees from \eref{Equ for C1}--\eref{Equ for C3} that $P(\psi_0,L,\Phi,\epsilon,t) - P_{\infty}(\psi_0,\Phi)$ behaves as $\epsilon^2 / a^4$ if $\cos(\Phi) = 1$ and $\epsilon^2/b^4$ otherwise. Looking at \Fig\ref{Fig3}, one deduces that at large times $P(\psi_0,L,\Phi,\epsilon,t) - P_{\infty}(\psi_0,\Phi)$ decreases with $\epsilon$ if $\cos(\Phi) = 1$ and increases otherwise. 
In other words, in the long run the particle escapes faster as one increases $\epsilon$ if $\cos(\Phi) = 1$ and escapes slower otherwise. This slowing effect when $\cos(\Phi) \not= 1$ at large times is rather peculiar since in the short time regime, 
some numerical results show (see \Fig\ref{Fig10}) that the particle may escape faster as $\epsilon$ increases. 

\newpage

\section{The Bound States}\label{Numerical Results}

In order to apply our general results, let us consider the following situation: the system is prepared such that at time $t=0$ no magnetic flux crosses the loop and the particle is in a bound state $\varphi_{n} \in D(H)$, where 
\begin{equation}\label{Bound states}
\varphi_{n}(x) = \sqrt{\frac{2}{L}} \sin\left(\frac{2 \pi n}{L} x\right) \chi_{[0,L)}(x)~, \hspace{5mm} n = 1,2\dots~.
\end{equation}
Since the effect of the length's loop and of the coupling parameter was thoroughly discussed in \Section~\ref{Discussion}, let us fix $L$ and $\epsilon$ and discuss the $\Phi$-dependence of $P(\varphi_{n},L,\Phi,\epsilon,t)$. If at all later times there is still no magnetic flux inside the loop $(\Phi=0)$, then all the bound states $\varphi_n$ will lead naturally to $P(\varphi_{n},L,\Phi,\epsilon,t)=1$ at all times, \ie the particle will remain forever in the loop. On the contrary, 
if some constant magnetic field is applied $(\Phi\not=0)$, then the states $\varphi_n$ will no longer be bound states of the Hamiltonian and consequently the particle may escape from the loop. Using \eref{Pinfinity}, one finds the following asymptotic values: 
\begin{equation}\label{Pasymp}
\lim_{t \rightarrow \infty} P(\varphi_{n},L,\Phi,\epsilon,t) = P_{\infty}(\varphi_{n},\Phi) = \left\{\begin{array}{ll} 1 & \mbox{if } \Phi = 0 \\ 0 & \mbox{if } |\cos(\Phi)| \not= 1    \\ 1/4 & \mbox{if } |\Phi| = 2 n \pi \\ 1/2 & \mbox{otherwise} \end{array}\right.~.
\end{equation}

\begin{figure}[htbp]
\centerline{\includegraphics[width=10cm]{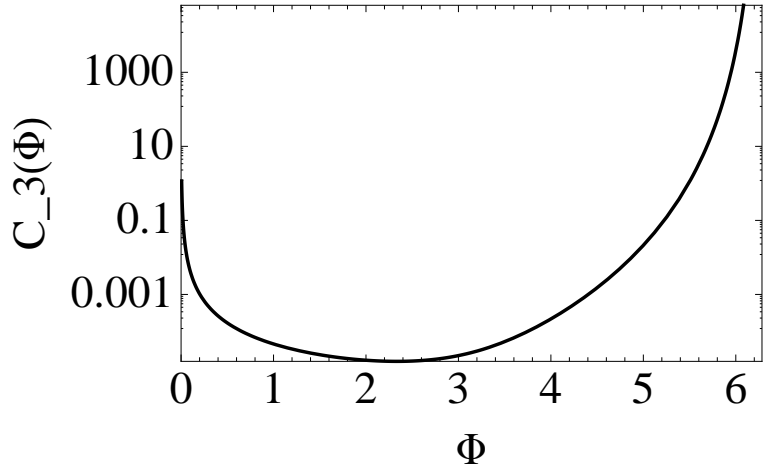}}
\caption{Log-lin plot of $C_3(\varphi_{n},L,\Phi,\epsilon)$ in \eref{Equ C3 exp} for $n=1$, $L=1$, $\epsilon=1/2$ and $\Phi \in (0,2\pi)$.}\label{Fig4}
\end{figure}

\begin{figure}[htbp]
\centerline{\includegraphics[width=9.5cm]{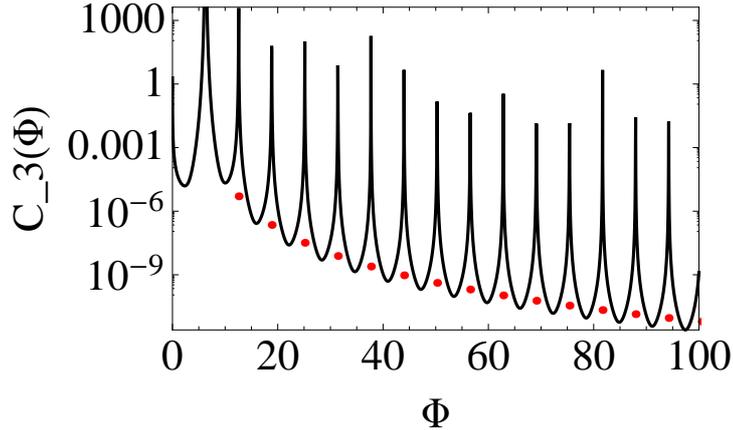}}
\caption{Log-lin plot of $C_3(\varphi_{n},L,\Phi,\epsilon)$ for $n=1$, $L=1$, $\epsilon=1/2$ and $\Phi \in [0,100]$. When $\Phi=0$, $C_3(\varphi_{n=1},L=1,\Phi,\epsilon=1/2)=0$. When $\Phi=2\pi$, $C_1(\varphi_{n=1},L=1,\Phi,\epsilon=1/2)\not=0$ so $C_3(\varphi_{n=1},L=1,\Phi,\epsilon=1/2)$ is not plotted. When $\Phi = 4\pi, 6\pi, \dots$, we use \eref{Coeff C3 two} to obtain the red dots. All the other values are obtained with \eref{Equ C3 exp}.}\label{Fig5}
\end{figure}

As explained in \Section~\ref{Discussion}, $P(\varphi_{n},L,\Phi,\epsilon,t)$ is invariant under $\Phi \mapsto -\Phi$ since $\varphi_n$ are odd functions with respect to $L/2$, so it is sufficient to discuss the case $\Phi \geq 0$. If $\Phi=0$, then $P(\varphi_{n},L,\Phi,\epsilon,t)=1$ at all times. If $\Phi \in (0,2\pi)$, then by using \eref{Pasymp} and \eref{Equ for C3} one finds $P(\varphi_{n},L,\Phi,\epsilon,t) = \delta_{\Phi, \pi}/2 + C_3(\varphi_{n},L,\Phi,\epsilon)/t^3 + \mathcal{O}(1/t^4)$, where
\begin{equation}\label{Equ C3 exp}
C_3(\varphi_{n},L,\Phi,\epsilon) = \frac{2 n^2 \epsilon^2 \Phi^2 L^6 \pi [2+\cos(\Phi)]}{3 b^4 [1-\cos(\Phi)]^2 (\Phi - 2n\pi)^4  (\Phi + 2n\pi)^4}~.
\end{equation}
A simple analysis of \eref{Equ C3 exp} reveals that $C_3(\varphi_{n},L,\Phi,\epsilon)$ is always positive and diverges at the excluded values $\cos(\Phi)=1$. Furthermore, we see in \Fig\ref{Fig4} that $C_3(\varphi_{n},L,\Phi,\epsilon)$ first decreases and then increases with $\Phi$. Note also that $C_3(\varphi_{n},L,\Phi,\epsilon)$ decreases with $n$, showing that in the long run more energetic bound states escape faster from the loop.  If $\Phi = 2\pi$ or more generally if $\cos(\Phi)=1$, then by using \eref{Equ for C1}--\eref{Equ C3 when C1 is zero} one finds $P(\varphi_{n},L,\Phi,\epsilon,t) = P_{\infty}(\varphi_{n},\Phi) + C_1(\varphi_{n},L,\Phi,\epsilon)/t + C_2(\varphi_{n},L,\Phi,\epsilon)/t^2 + C_3(\varphi_{n},L,\Phi,\epsilon)/t^3 + \mathcal{O}(1/t^4)$, where 
\begin{equation}\label{Equ C1}
C_1(\varphi_{n},L,\Phi,\epsilon) = \left\{\begin{array}{ll} L^2 \epsilon^2/(32 \pi a^4) & \mbox{if } |\Phi|=2n\pi \\ 0 & \mbox{otherwise}  \end{array}\right.~.
\end{equation}
If $C_1(\varphi_{n},L,\Phi,\epsilon)=0$ (\ie $|\Phi|\not=2n\pi$), then $C_2(\varphi_{n},L,\Phi,\epsilon)=0$ and
\begin{equation}\label{Coeff C3 two}
C_3(\varphi_{n},L,\Phi,\epsilon) = \frac{n^2 \epsilon^2 \Phi^2 L^6 \pi}{2 a^4 (\Phi - 2n\pi)^4  (\Phi + 2n\pi)^4}~.
\end{equation}
Note that, as in \eref{Equ C3 exp}, the coefficient $C_3(\varphi_{n},L,\Phi,\epsilon)$ in \eref{Coeff C3 two} decreases with $n$. Note also that $C_3(\varphi_{n},L,\Phi,\epsilon)$ decreases with $\Phi$, as one may easily see by looking at the red dots in \Fig\ref{Fig5}. Finally, we see in \Fig\ref{Fig5} that by increasing further the magnetic flux $\Phi$ we obtain a series of similar patterns in each range $[2\ell \pi, 2(\ell+1)\pi]$, $\ell = 0, 1, 2, \dots$

To obtain a global picture in time, we have computed $P(\varphi_{n},L,\Phi,\epsilon,t)$ numerically. For this, we wrote \eref{propagator details1} as
\begin{equation}
K_1(x,y,t) = \int_{-\infty}^{\infty} G(k;x,y,t) dk = 2 \int_{0}^{\infty} G^{\rm s}(k;x,y,t) dk~,
\end{equation}
where
\begin{equation}
G^{\rm s}(k;x,y,t) = \frac{1}{2} [G(k;x,y,t) + G(-k;x,y,t)]~,
\end{equation}
so that
\begin{equation}\label{Global Num}
P(\varphi_n,L,\Phi,\epsilon,t) = 4 \int_0^\infty  \int_0^\infty \left[\int_0^L \int_0^L \int_0^L G^{\rm s}(k;x,y,t) \overline{G^{\rm s}(p;x,z,t)} \psi_0(y) \overline{\psi_0(z)} dx  dy dz \right] dk  dp~.
\end{equation}
Then, we integrated analytically the expression inside the square brackets and then computed numerically (with Matlab) the so obtained expression with respect to $k$ and $p$ over the finite domain $[0,c]\times[0,c]$. The constant $c$ was chosen such that at very short times $P(\varphi_{n},L,\Phi,\epsilon,t)$ was close to one and no appreciable changes occurred in $P(\varphi_{n},L,\Phi,\epsilon,t)$ by increasing $c$ further. We also checked that the curve for ($\epsilon = 1/2$, $\Phi = \pi/2$), for which we know the propagator exactly [see \eref{Exact propagator 1}], was accurately reproduced. In all the curves having reached the asymptotic regime, the long time values are in very good agreement with the ones given by the relations \eref{Equ C3 exp}--\eref{Coeff C3 two}.

\begin{figure}[htbp]
\centerline{\includegraphics[width=12cm]{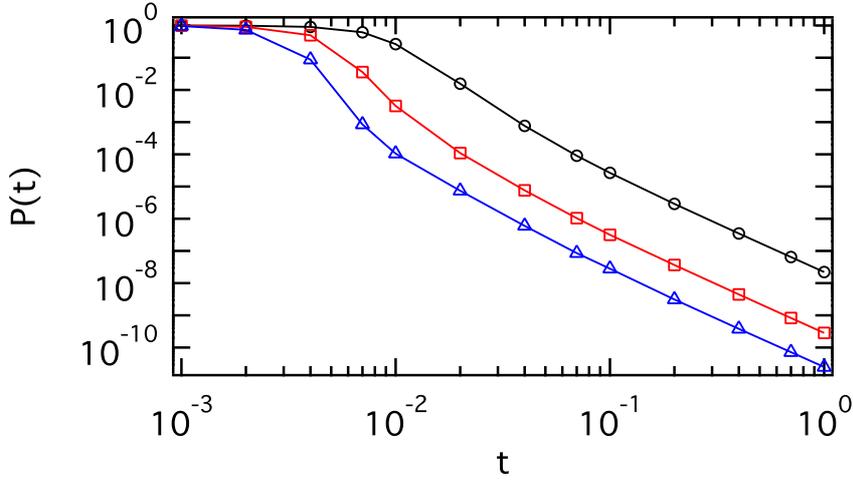}}
\caption{Log-log plot of $P(\varphi_n,L,\Phi,\epsilon,t)$ for $L=1$, $\epsilon=1/2$, $\Phi=\pi/2$ and $n=1$ (black circle), $n=2$ (red square) and $n=3$ (blue triangle).}\label{Fig6}
\end{figure}

{\bf $\boldsymbol{n}$-dependence:} One sees in \Fig\ref{Fig6} that $P(\varphi_{n},L,\Phi,\epsilon,t)$ decreases with $n$ at all times and that at large times $P(\varphi_n,L,\Phi,\epsilon,t) \sim C(\varphi_{n},L,\Phi,\epsilon)/t^3$ in good agreement with \eref{Equ C3 exp}.

\begin{figure}[htbp]
\centerline{\includegraphics[width=12cm]{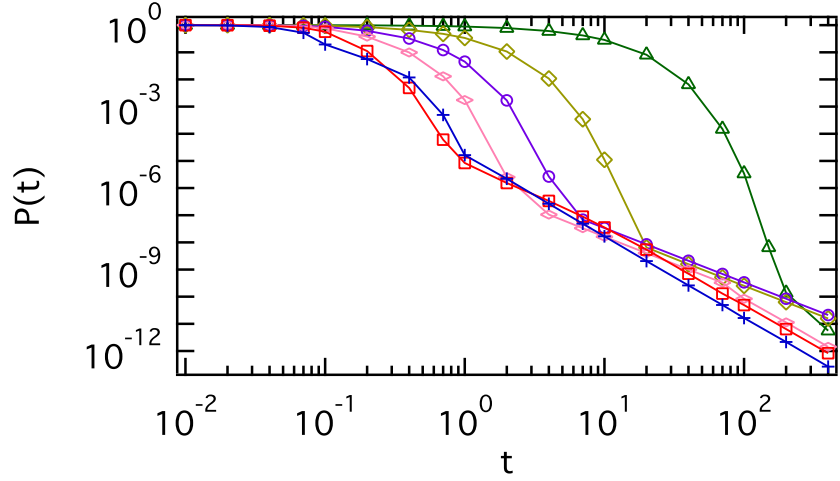}}
\caption{Log-log plot of $P(\varphi_n,L,\Phi,\epsilon,t)$ for $n=1$, $L=1$, $\epsilon=1/2$ and $\Phi=0.1$ (green triangle), $\Phi=0.3$ (yellow diamond), $\Phi=0.5$ (purple circle), $\Phi=0.7$ (pink lozenge), $\Phi=1$ (red square) and $\Phi=2$ (blue cross).}\label{Fig7}
\end{figure}

\begin{figure}[htbp]
\centerline{\includegraphics[width=12cm]{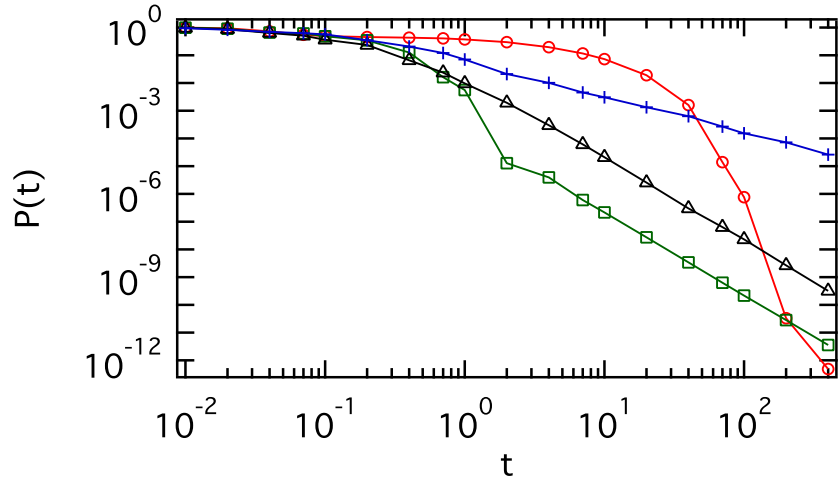}}
\caption{Log-log plot of $P(\varphi_n,L,\Phi,\epsilon,t)$ for $n=1$, $L=1$ $\epsilon=1/2$ and $\Phi=3$ (red circle), $\Phi=4$ (green square), $\Phi=5$ (black triangle) and $\Phi=6$ (blue cross).}\label{Fig8}
\end{figure}

\begin{figure}[htbp]
\centerline{\includegraphics[width=12cm]{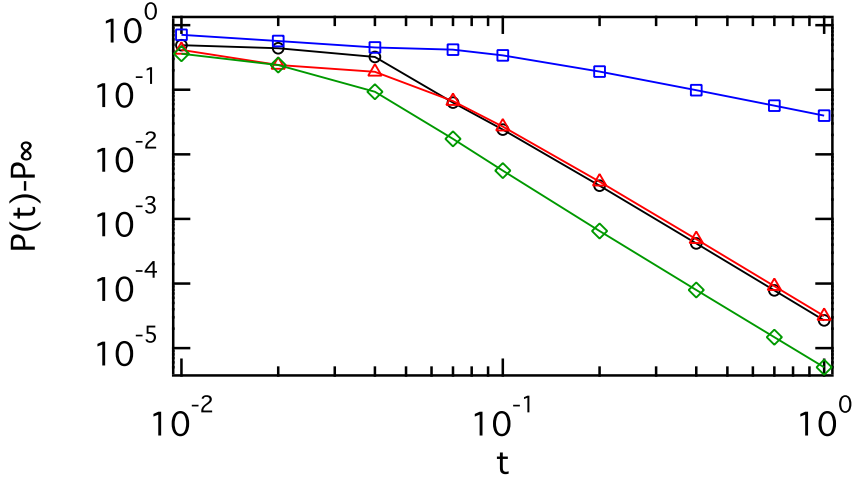}}
\caption{Log-log plot of $P(\varphi_n,L,\Phi,\epsilon,t)-P_{\infty}(\varphi_n,\Phi)$ for $n=1$, $L=1$, $\epsilon=1/2$ and $\Phi=\pi$ (black circle), $\Phi=2 \pi$ (blue square), $\Phi=3 \pi$ (red triangle) and $\Phi=4 \pi$ (green diamond).}\label{Fig9}
\end{figure}

{\bf $\boldsymbol{\Phi}$-dependence:} The $\Phi$-dependence of $P(\varphi_n,L,\Phi,\epsilon,t)$ is represented in \Figs\ref{Fig7}--\ref{Fig9}. In the curves having reached the asymptotic regime, one sees that $P(t) \sim C(\varphi_{n},L,\Phi,\epsilon)/t^3$ at large times, 
where the constant $C(\varphi_{n},L,\Phi,\epsilon)$ decreases with $\Phi$ when $\Phi \in (0,2.33)$ (see \Fig\ref{Fig7}) and increases when $\Phi \in (2.33,2\pi)$  (see \Fig\ref{Fig8}), 
as expected from \Fig\ref{Fig4}. Note however that there may be several crossing among the curves. Note that similar features were obtained in a circular dielectric cavity containing classical waves \cite{Ryu}. In \Fig\ref{Fig9}, one sees that $\Phi = 2\pi$ leads to $P(\varphi_n,L,\Phi,\epsilon,t) \sim C(\varphi_{n},L,\Phi,\epsilon)/t$ as expected from \eref{Equ C1}, while $\Phi = \pi, 3\pi, 4\pi$ lead to $P(\varphi_n,L,\Phi,\epsilon,t) \sim C(\varphi_{n},L,\Phi,\epsilon)/t^3$ as expected from \eref{Equ C3 exp} and \eref{Coeff C3 two}.

\begin{figure}[htbp]
\centerline{\includegraphics[width=12cm]{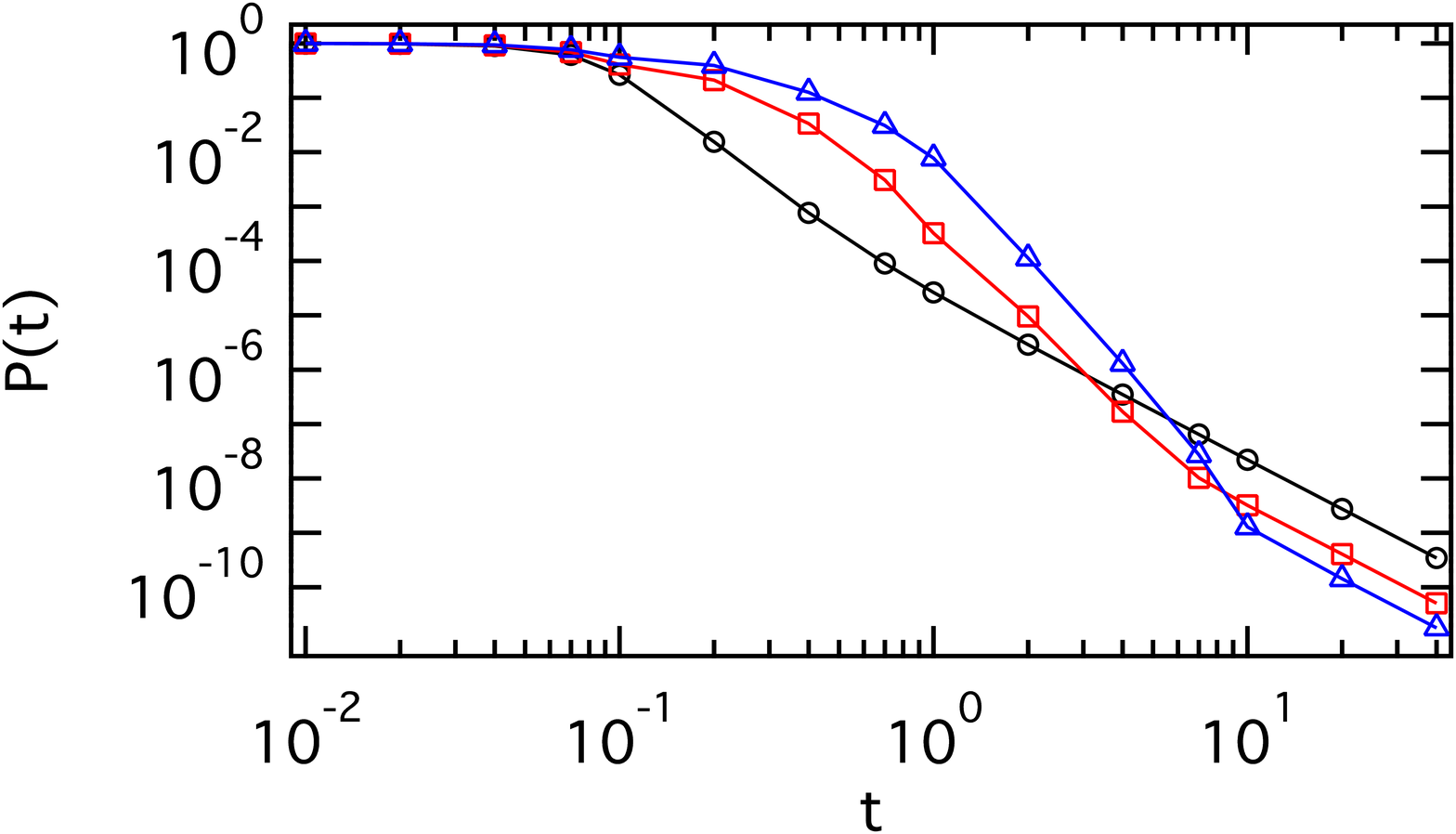}}
\caption{Log-log plot of $P(\varphi_n,L,\Phi,\epsilon,t)$ for $n=1$, $L=1$, $\Phi=\pi/2$ and $\epsilon=0.3$ (blue triangle), $\epsilon=0.4$ (red square) and $\epsilon=0.5$ (black circle).}\label{Fig10}
\end{figure}

{\bf $\boldsymbol{\epsilon}$-dependence:} In \Fig\ref{Fig10}, one sees that $P(\varphi_n,L,\Phi,\epsilon,t) \sim C(\varphi_{n},L,\Phi,\epsilon)/t^3$ at large times and that the constant $C(\varphi_{n},L,\Phi,\epsilon)$ increases with $\epsilon$. On the other hand, note that $P(\varphi_n,L,\Phi,\epsilon,t)$ decreases with $\epsilon$ at short times and that the curves cross each other.

{\bf Exponential decay:}  Plotting the curves in \Figs\ref{Fig6}--\ref{Fig10} on a log-lin plot (not shown) reveals that on some intermediate time scale (not very small and not very large), the nonescape probability decays approximately as an exponential, in agreement with \cite{Ballentine}.

\newpage

$\ $

\vspace{40mm}

\section{The Interacting case}\label{The Interacting case}

So far we have solely discussed the free evolution. To have an idea about the interacting evolution, let us consider the case $\Phi=0$ and $\epsilon=4/9$, where the Hamiltonian of the particle is given by
\begin{equation}\label{Hamiltonian Int}
H = H_0 + V(Q)~.  
\end{equation}
Here $H_0=-\frac{d^2}{dx^2}$ is the free Hamiltonian and $V$ is the potential which for simplicity is supposed to be bounded, so that the Hamiltonian $H$ is self-adjoint in the domain $D(H)$ specified by the conditions \eref{Condition 1}--\eref{Condition 3} with $\epsilon=4/9$.

Obviously, in this case the time-dependent Schr\"odinger equation \eref{Schrodinger equ} cannot be solved analytically in general and it also seems extremely difficult to compute analytically the large time behaviour of the nonescape probability, so we shall consider a fully numerical approach. For this, let us consider a long but finite lead of length $\ell > L$ and set $\psi(L+\ell)=0$. The assumption $\ell>L$ is natural in our case and also simplifies the remaining discussion. One can then check that $H$ is self-adjoint in 
\begin{equation}
 D(H) \hspace{-1mm} = \hspace{-1mm} \{ \psi \in \mathcal{H} \ | \ H \psi \in \mathcal{H}  \mbox{ and }  \eref{Condition 1p}-\eref{Condition 3p} \mbox{ are satisfied}\}
\end{equation}
where
\begin{eqnarray}
 \psi(0_+)=\psi(L_-)=\psi(L_+) ~, \label{Condition 1p}\\
 \psi'(L_-)=\psi'(0_+) + \psi'(L_+) ~, \label{Condition 2p}\\
\psi(L+\ell)=0 \label{Condition 3p}~.
\end{eqnarray}

Clearly, for our numerical results to remain accurate for a long time the unitarity of the time-evolution is crucial. A convenient way to implement numerically the boundary conditions  \eref{Condition 1p}--\eref{Condition 3p}, while preserving unitarity, is to use the pseudo-spectral method presented in Appendix~D. This method requires the knowledge of the spectrum and bound states of the free Hamiltonian $H_0$, which are worked out in Appendix~E. We believe that this section and the Appendices D and E present, in particular, an interesting and highly non-trivial application of the pseudo-spectral method.

\begin{figure}[htbp]
\centerline{\includegraphics[width=12cm]{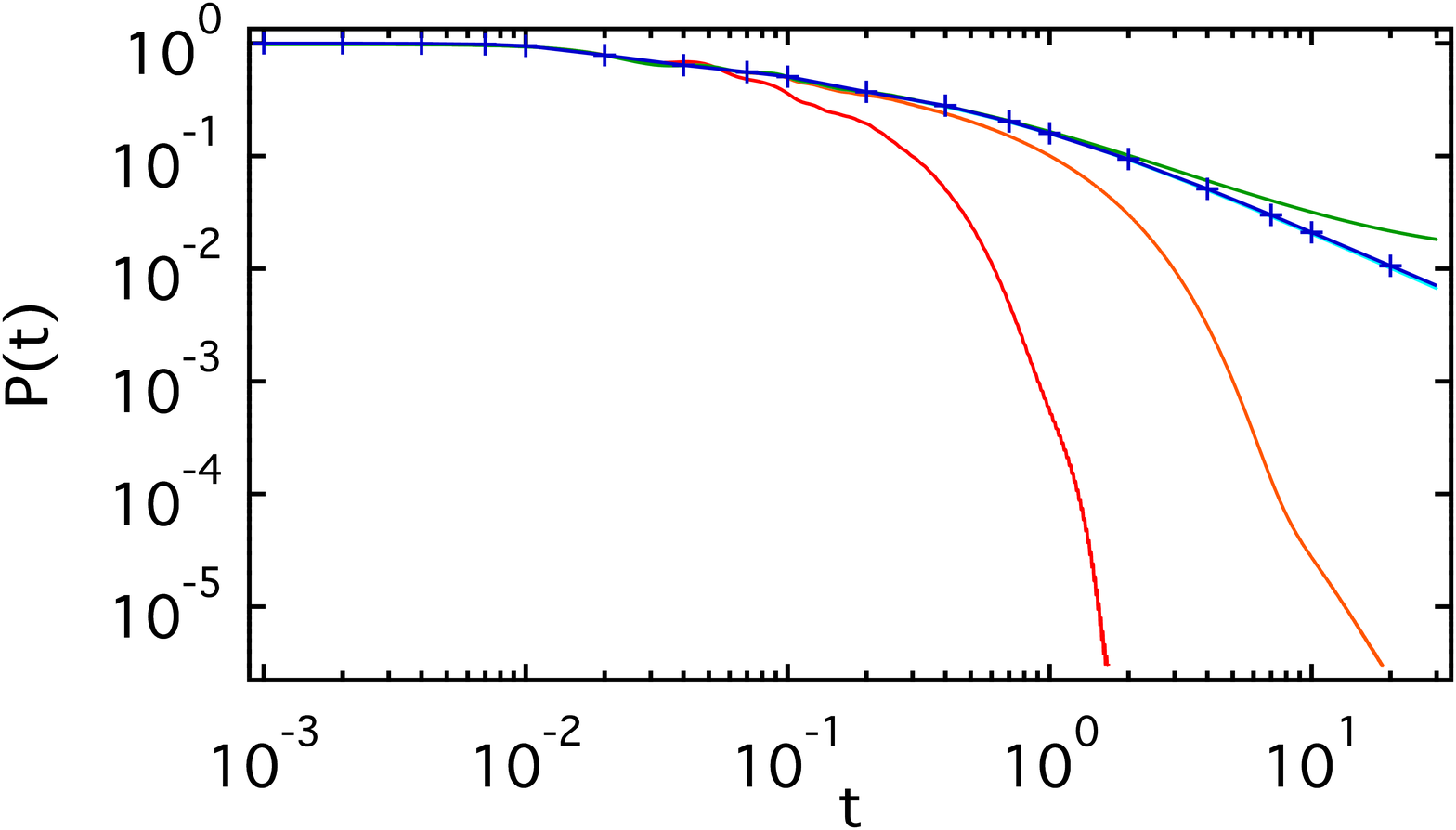}}
\caption{The nonescape probability $P(\psi_0,L,\ell,\lambda,t)$ for a soft-core coulomb potential by using the pseudo-spectral method with $L=1$, $\ell = 500$, $\Delta x = 1/80$, $\Delta t = 10^{-3}$. From top to bottom: $\lambda = -0.1$ (green), $\lambda = 0$ (light blue), $\lambda = 0.1$ (orange) and $\lambda = 1$ (red). The dark blue cross curve (close to the $\lambda=0$ curve) corresponds to the free case ($V=0$) with $\ell = \infty$ and is obtained by using the expression \eref{Global Num}.}\label{Fig11}
\end{figure}

Let us consider the following initial state:
\begin{equation}\label{Initial State No Bound}
\psi_0(x) = \sqrt{\frac{8}{3 L}} \sin^2\left(\frac{2\pi}{L}x\right) \chi_{[0,L]}(x)~.
\end{equation}
It is easy to check that $\psi_0$ is normalized ($||\psi_0||=1$) and belongs to $D(H)$. Note that $\psi_0$ is an even function in $L^2([0,L])$ and thus it is orthogonal to all bound states of the loop. Therefore, in the infinite case $\ell = \infty$ with no potential ($V=0$), the nonescape probability will vanish at large times and the asymptotic decay can be computed by using \eref{Equ for C1}:  
\begin{equation}\label{Asymptotic Potential}
P_{\rm free}(\psi_0,L,\ell=\infty,t) = \frac{2 L^2}{3 \pi} \ \frac{1}{t} + \mathcal{O}\left(\frac{1}{t^2}\right)~.
\end{equation} 
Numerically, the infinite situation ($\ell = \infty$) is obtained by using the expression \eref{Global Num} and gives the dark blue cross curve in Fig.~\ref{Fig11}. Its asymptotic values are in very good agreement with formula \eref{Asymptotic Potential}. On the other hand, the finite case ($\ell < \infty$) is obtained by using the pseudo-spectral method and leads to the light blue curve in Fig.~\ref{Fig11}. As one clearly sees, the two methods are in very good agreement. 

Imagine now that a fictive charged particle is fixed at the point $x=L/2$, \ie at the leftmost part of the loop, and interacts with our genuine charged particle via a soft-core coulomb force. Then, a simple calculation leads to the following potential ($0 \leq x \leq L+\ell$): 
\begin{equation}
V_\lambda(x) = \frac{\lambda}{\sqrt{d(x)^2 + 10^{-4}}}~, \hspace{2mm} d(x) = \left\{\begin{array}{ll} \sqrt{2} R \sqrt{1-\cos\left(\pi - \frac{x}{R}\right)} & \mbox{if }  x \leq L \\ 2R+x-L & \mbox{if } x > L \end{array}\right.~, \hspace{2mm} R=\frac{L}{2\pi}~.
\end{equation}
When $\lambda > 0$, the force is repulsive, while $\lambda < 0$ corresponds to an attractive force. The numerical results are presented in \Fig\ref{Fig11}. As expected, we see that $P(\psi_0,L,\ell,\lambda,t)$ decays faster than the free case if $\lambda > 0$ and slower if $\lambda < 0$. Note also that globally the curves look similar to the ones obtained in the free cases presented in the previous section. In particular, plotting these curves on a log-lin plot (not shown) reveals the approximate exponential decay on some intermediate time scale. Although our numerical results do not show any asymptotic power law decay when $\lambda \not = 0$, such a regime may well be present at larger times. To reach larger times, however, one has to increase the length of the lead $\ell$ (because the particle gets reflected at the end of the lead and comes back into the loop causing a bump in the curves after some time) but this increases the computation time very quickly.

\section{Concluding Remarks}\label{Conclusion}

Although more energetic bound states typically escape faster from the loop, we found that different bound states may actually decay with different power laws and consequently it may be possible to filter one bound state from a coherent superposition by applying the appropriate magnetic flux $\Phi$ through the loop. Indeed, suppose that the system is prepared such that at time $t=0$ no magnetic flux crosses the loop and the particle is in the state 
\begin{equation}
\psi_0 = \sum_{n=1}^\infty c_n \varphi_{n}~,
\end{equation}
satisfying $||\psi_0||=1$, where $\varphi_{n}$ are the bound states \eref{Bound states}. Then, if one wants to \emph{filter} the bound state $\varphi_{m}$, 
one may apply a magnetic flux $\Phi = 2m\pi$ through the loop for a sufficiently long time, so that the state $\varphi_{m}$ decays as $1/t$ while all the others decay as $1/t^3$ [see \eref{Equ C1}--\eref{Coeff C3 two}]. 

The short-time behaviour of $P(\psi_0,L,\Phi,\epsilon,t)$ is also interesting and can be conveniently written as $P(\psi_0,L,\Phi,\epsilon,t)=1-||\chi_{[L,\infty)}(Q)e^{-iHt}\psi_0||^2$. By computing the propagator $K(x,y,t)$ for $x>L$ and $y<L$, one then founds that $P(\psi_0,L,\Phi,\epsilon,t)=1-D_2(\psi_0,L,\Phi,\epsilon) t^2 + \mathcal{O}(t^3)$ as $t \rightarrow 0$. In general, it seems that the constant $D_2(\psi_0,L,\Phi,\epsilon)$ cannot be expressed in a simple form and thus its behaviour would require a refined numerical investigation.

To make more apparent the exponential decay of the nonescape probability at intermediate times, while keeping the power law decay explicit at large times, it would be interesting to write $P(\psi_0,L,\Phi,\epsilon,t)$ as $P_\infty(\psi_0,\Phi) + A(\psi_0,L,\Phi,\epsilon) \exp[-\alpha(\psi_0,L,\Phi,\epsilon) t] + \sum_{k=1}^\infty B_k(\psi_0,L,\Phi,\epsilon)/t^k$ and to determine the unknown variables $A$, $\alpha$ and $B_k$. One may also analyse the transition times between the exponential regime and the power law regime with respect to the parameters $\epsilon$ and $\Phi$, and also $\lambda$ if there is such a transition.

\section*{Acknowledgements}

The author thanks T. Taniguchi for suggesting him this problem and for stimulating discussions. He also thanks W. O. Amrein, M. B\"uttiker, J.-P. Eckmann,  P. Exner, J.~Jacquet, Ph. Jacquod and C.-A. Pillet for helpful comments and E. Hairer and C. Lubich for useful remarks related to the numerical simulations. This work was partially supported by the Japan Society for the Promotion of Science.

\renewcommand{\theequation}{A\arabic{equation}}
\setcounter{equation}{0}
\section*{Appendix A: Properties of the Hamiltonian $\boldsymbol{H}$}

Here we show that the Hamiltonian $H$ is self-adjoint in $D(H)$ given in \eref{Domain of H}. First of all, it is clear that $D(H)$ is dense in $\mathcal{H}$, so that the adjoint $H^*$ of $H$ is well defined. Let us recall that an operator $\{H,D(H)\}$ is self-adjoint if $D(H^*)=D(H)$ and $H^* \psi = H \psi$ for all $\psi \in D(H)$, where a vector $f \in \mathcal{H}$ belongs to $D(H^*)$ 
if there is a vector $f^* \in \mathcal{H}$ such that $\langle f | H g \rangle = \langle f^* | g \rangle$ for all $g \in D(H)$, and for each $f \in D(H^*)$ one sets $H^* f = f^*$.

Let $f \in \mathcal{H}$ be a vector such that $H f \in \mathcal{H}$ and let $g \in D(H)$. Then, 
\begin{equation}
\langle f | H g \rangle = \langle H f | g \rangle + \mbox{BT}~,
\end{equation}
where 
\begin{eqnarray*}
\mbox{BT} &=&  2i \frac{\Phi}{L} g(0_+)  \left[\overline{f(L_-)}  - \overline{f(0_+)}\right] - \frac{g(0_+)}{b} \left\{\sqrt{\epsilon} \overline{f'(L_+)} +b \left[\overline{f'(0_+)} - \overline{f'(L_-)}\right]\right\}  \\
&& +  \frac{g'(0_+)}{\sqrt{\epsilon}} \left[\sqrt{\epsilon} \overline{f(0_+)} - b \overline{f(L_+)}\right] - \frac{g'(L_-)}{\sqrt{\epsilon}} \left[\sqrt{\epsilon} \overline{f(L_-)} - b \overline{f(L_+)}\right]~.
\end{eqnarray*}
Here we have used the fact that $g \in D(H)$. We see that $\mbox{BT}$ vanishes for all $g \in D(H)$ if and only if $f$ also satisfies the boundary conditions \eref{Condition 1}-\eref{Condition 3}. 
This shows that $D(H^*)=D(H)$ and $H^* f = H f$ for all $f \in D(H)$, \ie $H$ is self-adjoint.

Let us now show that the spectrum $\sigma(H)$ of $H$ is $[0,\infty)$. Since $H$ is self-adjoint, it follows that $\sigma(H) \subset \real$. Writing $H=P_\Phi^2$, with $P_\Phi = -i\frac{d}{dx} - \frac{\Phi}{L} \chi_{[0,L]}(Q)$, 
one easily shows that $\langle \varphi | H \varphi \rangle = ||P_\Phi \varphi||^2 \geq 0$ for all $\varphi \in D(H)$. 
This means that $H$ is a positive operator and thus $\sigma(H) \subset [0,\infty)$. Let $k > 0$ be fixed, then a (particular) solution of  $H \varphi = k^2 \varphi$ is
\begin{equation}
\varphi(x) = e^{i k (x-L)} \chi_{[L,\infty)}(x)~.
\end{equation}
Take $\psi \in C^\infty_0(\real)$ satisfying $\psi(0)=\psi'(0)=0$ and $\int_{0}^\infty |\psi(y)|^2 dy = 1$, and set
\begin{equation}
\psi_{n}(x) =  \frac{1}{\sqrt{n}} \psi\left(\frac{x-L}{n}\right) \varphi(x)~, \hspace{2mm} \mbox{where} \hspace{2mm} x > 0  \hspace{2mm} \mbox{and} \hspace{2mm} n = 1,2,\dots
\end{equation}
It is easy to check that $\psi_n \in D(H)$ and $||\psi_n||=1$ for all $n$, and that $||H \psi_n - k^2 \psi_n|| \rightarrow 0$ as $n \rightarrow \infty$. This shows that $E=k^2 > 0$ belongs to the spectrum of $H$ (see Prop.~4.20 in \cite{Amrein} or Th.~2.2.1 in \cite{Schechter}). Since $k>0$ is arbitrary and the spectrum of $H$ is a closed set, one can conclude that $\sigma(H) = [0,\infty)$. 

\renewcommand{\theequation}{B\arabic{equation}}
\setcounter{equation}{0}
\section*{Appendix B: The coefficients $\boldsymbol{A}$, $\boldsymbol{B}$, $\boldsymbol{C}$ and $\boldsymbol{D}$}

The denominator $\mbox{DEN}(k)$ of the coefficients \eref{Coeff A}--\eref{Coeff D} vanish if and only if 
\begin{equation}
a^2 \sin^2(kL) + b^2 [\cos(kL) - \cos(\Phi)]^2 = 0~.
\end{equation}
Since $a$ and $b$ are always non-zero (remember that $\epsilon = 0$ is excluded) the above equality is satisfied if and only if 
\begin{equation}
\sin(kL) = 0 \hspace{5mm} \mbox{and} \hspace{5mm} \cos(kL) = \cos(\Phi)~.
\end{equation}
The first relation is verified if only if $kL = n \pi$, for some $n \in \mathbb{Z}$, and consequently the second relation reads $\cos(\Phi) = (-1)^n$. 
This shows that the coefficients $A$, $B$, $C$ and $D$ may contain singularities only if $\cos(\Phi) = 1$ or $\cos(\Phi) = -1$. We shall now discuss these two cases separately.\\ \\
(i) Case $\cos(\Phi) = 1$ or $\Phi = 2 n \pi$, with $n \in \mathbb{Z}$: Using trigonometric relations, one may write
\begin{equation}
A(k) = -\overline{B(k)} = -\sqrt{\frac{\epsilon}{8 \pi}} \frac{\sqrt{1 - \cos(kL)} + i \ \mbox{sign}[\sin(kL)] \sqrt{1 + \cos(kL)}}{\left\{a^2 [1 + \cos(kL)] + b^2 [1 - \cos(kL)]\right\}^{1/2}}~.
\end{equation}
Since the denominator of this expression is always non-zero, it follows that $A$ and $B$ are well defined in $\real$. Next, one may write
\begin{equation}\label{Relation C and D}
C(k) = -\overline{D(k)} = -\frac{1}{2 b \sqrt{\epsilon}} [2 (b-\epsilon) A(k)- 2 b B(k)]~,
\end{equation}
showing that $C$ and $D$ are also well defined in $\real$. Furthermore, one has
\begin{eqnarray}
f_1(k) &=& \frac{\epsilon}{4\pi \{a^2 [1 + \cos(kL)] + b^2 [1 - \cos(kL)]\}}~,\label{f1 cos1}\\
f_2(k) &=& \frac{\epsilon e^{-ikL}}{4\pi \{a^2 [1 + \cos(kL)] + b^2 [1 - \cos(kL)]\}}~.\label{f2 cos1}
\end{eqnarray}
(ii) Case $\cos(\Phi) = -1$ or $\Phi = (2n+1) \pi$, with $n \in \mathbb{Z}$: In this case, the relation \eref{Relation C and D} still holds with 
\begin{equation}
A(k) = -\overline{B(k)} = -\sqrt{\frac{\epsilon}{8 \pi}} \frac{\sqrt{1 + \cos(kL)} - i \ \mbox{sign}[\sin(kL)] \sqrt{1 - \cos(kL)}}{\left\{a^2 [1 - \cos(kL)] + b^2 [1 + \cos(kL)]\right\}^{1/2}}~.
\end{equation}
Therefore, the coefficients are also well defined in this case and we have
\begin{eqnarray}
f_1(k) &=& \frac{\epsilon}{4\pi \{a^2 [1 - \cos(kL)] + b^2 [1 + \cos(kL)]\}}~,\label{f1 cosm1}\\
f_2(k) &=& \frac{-\epsilon e^{-ikL}}{4\pi \{a^2 [1 - \cos(kL)] + b^2 [1 + \cos(kL)]\}}~.\label{f2 cosm1}
\end{eqnarray}

\renewcommand{\theequation}{C\arabic{equation}}
\setcounter{equation}{0}
\section*{Appendix C: The identity \eref{Identity Gaussian Integral}}

Here we show the identity \eref{Identity Gaussian Integral}. From the equality
\begin{equation}
\int_{-\infty}^{\infty} e^{-ak^2+bk} dk = \sqrt{\frac{\pi}{a}} e^{\frac{b^2}{4a}}
\end{equation}
one can write
\begin{equation}
\int_{-\infty}^{\infty} k^n e^{-ak^2+bk} dk = \sqrt{\frac{\pi}{a}} \ \frac{d^n}{db^n} \left(e^{\frac{b^2}{4a}}\right)~.
\end{equation}
One then concludes by using the following formula for the Hermite polynomial:
\begin{equation}
H_n(x)= (-1)^n e^{x^2} \frac{d^n}{dx^n} \left(e^{-x^2}\right) = \sum_{\ell = 0}^{\lfloor \frac{n}{2} \rfloor} \frac{n! (-1)^{\ell}}{\ell! \ (n-2\ell)!} \ (2x)^{n-2\ell}~.
\end{equation}

\renewcommand{\theequation}{D\arabic{equation}}
\setcounter{equation}{0}
\section*{Appendix D: The Pseudo-Spectral Method}

In Appendix~E, we show that the free Hamiltonian $H_0$ has a pure point spectrum $\{E_n\}_{n=1}^\infty$ satisfying $0 < E_1 < E_2 < \cdots $ (so each eigenvalue is positive and has multiplicity 1), and that one can solve (explicitely) the free stationary Schr\"odindger equation 
\begin{equation}
H_0 \varphi_n = E_n \varphi_n~,
\end{equation}
with $\varphi_n \in D(H_0)=D(H)$ and $||\varphi_n|| = [\int_0^{L+\ell} |\varphi_n(x)|^2 dx]^{1/2} = 1$.  Let $\psi_0 \in D(H)$, $x \in [0,L+\ell]$ and $\Delta t > 0$, then 
\begin{equation}\label{Exact Sol}
\psi(x,\Delta t) = (e^{-iH\Delta t} \psi_0)(x)~.
\end{equation}
Let us consider the symmetric decomposition known as Strang splitting ($H=H_0 + V$):
\begin{equation}\label{Splitting}
e^{-iH\Delta t} = e^{-iV\Delta t/2} e^{-iH_0\Delta t} e^{-iV\Delta t/2} + \mathcal{O}(\Delta t^3)~.
\end{equation}
Since $\{\varphi_n\}_{n\geq 1}$ forms an orthonormal basis of the Hilbert space $\mathcal{H}$, one can write
\begin{equation}\label{Dev Base}
 e^{-iV\Delta t/2} \psi_0 = \sum_{n = 1}^{\infty} \langle \varphi_n | e^{-iV\Delta t/2} \psi_0 \rangle \varphi_n~.
\end{equation}
Substituting \eref{Splitting} and \eref{Dev Base} in \eref{Exact Sol} gives
\begin{equation}\label{Equ Num}
\psi(x,\Delta t) = \sum_{n = 1}^{\infty} e^{-iE_n \Delta t} e^{-iV(x)\Delta t/2} \varphi_n(x) \int_0^{L+\ell}  e^{-iV(y)\Delta t/2} \overline{\varphi_n(y)} \psi_0(y) dy + \mathcal{O}(\Delta t^3)~.
\end{equation}
Let $\Phi \in \mathcal{H}$, and let us define a generalized Fourier transform $\mathcal{F}$ as ($n = 1, 2, \dots$)
\begin{equation}\label{Eq1}
\hat{\Phi}(n)=\mathcal{F}(\Phi)(n) = \int_{0}^{L+\ell} \overline{\varphi_n(y)} \Phi(y) dy~.
\end{equation}
Since the family $\{\varphi_n\}_{n\geq 1}$ forms an orthonormal basis of $\mathcal{H}$, we have ($x \in [0,L+\ell]$):
\begin{equation}\label{Eq2}
\Phi(x)=\mathcal{F}^{-1}(\hat{\Phi})(x) = \sum_{m=1}^{\infty} \varphi_m(x) \hat{\Phi}(m)~.
\end{equation}
One may then rewrite \eref{Equ Num} as
\begin{equation}\label{Equ Num 2}
\psi(x,t+\Delta t) = e^{-iV(x)\Delta t/2}  \mathcal{F}^{-1} \left[e^{-iE_n \Delta t} \cdot \mathcal{F} \left(  e^{-iV(\cdot)\Delta t/2}  \psi(\cdot,t)\right) \right](x) + \mathcal{O}(\Delta t^3)~.
\end{equation}
This relation is a common way of presenting the pseudo-spectral method. Numerically, the Fourier transform and its inverse are then computed by using a very efficient algorithm known 
as the Fast Fourier Transform (FFT). As far as we know, there is however no FFT using the eigenvectors $\{\varphi_n\}_{n\geq 1}$ of our model, so we shall compute $\psi(x,t+\Delta t)$ directly. 

Let $\Delta x = (L+\ell)/N$, with $N \in \mathbb{N}$, be a small length difference, $x_k = k \ \Delta x$ the position of the $k$-th site, with $k=0,...,N$, $t_m = m \ \Delta t$, with $m = 0, 1, \dots$, and $\psi(\cdot,0) = \psi_0$. Then, we will use the following relation in the numerical computation:
\begin{equation}\label{Equ Num Final}
\psi(x_k,t_{m+1}) = \sum_{n = 1}^{\infty}  e^{-iE_n \Delta t} e^{-iV(x_k)\Delta t/2} \varphi_n(x_k) \int_0^{L+\ell}  e^{-iV(y)\Delta t/2} \overline{\varphi_n(y)} \psi(y,t_m) dy + \mathcal{O}(\Delta t^3)~.
\end{equation}
In the simulations, we consider only the first $2500$ terms in the infinite sum occurring in \eref{Equ Num Final}, compute the integral by using the composite Simpson rule and check that probability is conserved at all times.

\renewcommand{\theequation}{E\arabic{equation}}
\setcounter{equation}{0}
\section*{Appendix E: The free Hamiltonian $\boldsymbol{H_0}$}

The free Hamiltonian $H_0$ is self-adjoint and thus its spectrum $\sigma(H_0) \subset \real$. Writing $H_0=P^2$, with $P = -i\frac{d}{dx}$, 
one finds that $\langle \varphi | H_0 \varphi \rangle = ||P \varphi||^2 \geq 0$ for all $\varphi \in D(H)$ and thus $\sigma(H_0) \subset [0,\infty)$. 
The general solution of $H_0 \varphi_n = E_n \varphi_n$, with $E_n=k_n^2$ and $k_n \geq 0$, is
\begin{equation}
\varphi_n(x) = \left[A_n e^{ik_nx} +  B_n e^{-ik_nx}\right] \chi_{[0,L)}(x) + \left[C_n e^{ik_nx} +  D_n e^{-ik_nx}\right] \chi_{[L,L+\ell]}(x)~,
\end{equation}
where $\chi_{I}$ is the characteristic function of the interval $I$: $\chi_{I}(x) = 1$ if $x \in I$ and  $\chi_{I}(x) = 0$ otherwise. Solving \eref{Condition 1p}--\eref{Condition 3p}, one gets two types of solutions: First, we have the bound states of the loop ($C_n=D_n=0$ and $A_n=-B_n$):
\begin{equation}
\varphi^{\rm loop}_m(x) = \sqrt{\frac{2}{L}} \sin\left(k^{\rm loop}_m x\right) \chi_{[0,L)}(x)~, \hspace{4mm} k^{\rm loop}_m = \frac{2\pi m}{L}~,  \hspace{4mm}  m=1,2,\dots
\end{equation}
Second, assuming that $k_n \not= k^{\rm loop}_m$ for all $m=1,2,\dots$, we have
\begin{eqnarray}
B_n &=& - \frac{1-e^{ik_nL}}{1-e^{-ik_nL}} A_n~,\\
C_n &=& \frac{3 - 4 e^{-ik_nL}+e^{-2ik_nL}}{2(1-e^{-ik_nL})} A_n~,\\
D_n &=& -\frac{3-4e^{ik_nL}+e^{2ik_nL}}{2(1-e^{-ik_nL})} A_n~,\\
k_n &\not =& \frac{(2s+1) \pi}{2\ell}~, \hspace{3mm} s=0,1,2,\dots\\
2 \tan(k_n\ell)&=&\frac{\sin(k_nL)}{1-\cos(k_nL)}\label{Implicit}~.
\end{eqnarray}
Explicitly, we thus have
\begin{eqnarray}
\varphi^{\rm global}_n(x) &=& \frac{i A_n}{1-e^{-ik_nL}} \left\{2 \left[ \sin(k_nx)-\sin(k_n(x-L))\right] \chi_{[0,L)}(x) \right.\label{Global sol}\\
& & \hspace{-20mm}+ \left.  \left[3\sin(k_nx)-4\sin(k_n(x-L))+\sin(k_n(x-2L))\right] \chi_{[L,L+\ell]}(x)\right\}~,\nonumber
\end{eqnarray}
where the coefficient $A_n$ is determined by normalization: $\int_0^{L+\ell} |\varphi_n(x)|^2 dx = 1$. After some algebra, we found
\begin{eqnarray}
\hspace{-43mm}A_n &=& \left\{2\,L+\,\ell \left[ 5-3\,\cos \left( k_nL \right)
 \right] +2\,{\frac {\sin \left( k_nL \right)  \left[ 1-2\, \sin^{2}
 \left( k_n\ell\right)  \right] }{k_n}}\right.\nonumber\\
 && \hspace{39mm}\left.+\,{\frac {\sin \left( 
k_n\ell \right) \cos \left( k_n\ell \right)  \left[5\,\cos \left( k_nL \right) 
 -3\right] }{k_n}}\right\}^{-1/2}~.
\end{eqnarray}
It remains to solve \eref{Implicit}. As far as we can see, this relation cannot be solved explicitly. Since we shall sum up over the energies $E_n = k_n^2$, we actually only need to find out the non-negative solutions of \eref{Implicit}. Let $I_0 = [0,\frac{\pi}{2\ell}]$ and $I = [-\frac{\pi}{2\ell},\frac{\pi}{2\ell}]$, and let us look for solutions of the following form: 
\begin{equation}
k_n = n \frac{\pi}{\ell}+\delta_n~,
\end{equation}
where $n = 0, 1, 2, \dots$, $\delta_0 \in I_0$, $\delta_n \in I$ and such that $k_n \not = 2 \pi m/L$ and $k_n \not = (2m+1) \pi/(2\ell)$ for all $m=0,1,2,\dots$. Then, the relation \eref{Implicit} reads 
\begin{equation}
2 \tan(n \pi + \delta_n \ell) = \frac{\sin(n \frac{\pi L}{\ell} + \delta_n L)}{1-\cos(n \frac{\pi L}{\ell} + \delta_n L)}~.
\end{equation}
Since $\tan(n \pi + \delta_n \ell)=\tan(\delta_n \ell)$, one finds
\begin{equation}\label{Equ Cn}
\delta_n = \frac{1}{\ell} \arctan\left[\frac{\sin(n \frac{\pi L}{\ell} + \delta_n L)}{2 [1-\cos(n \frac{\pi L}{\ell} + \delta_n L)]} \right]~,
\end{equation}
where we consider the principal value of the arc-tangent function: $\arctan(\theta) \in (-\frac{\pi}{2},\frac{\pi}{2})$. Note that $\cos(n \frac{\pi L}{\ell} + \delta_n L) \not = 1$, since $k_n \not = 2 \pi m/L$ for all $m=0,1,2,\dots$, so the above expression is well defined. 
Nevertheless, it will be convenient to attribute values to this expression on these exceptional points. 

The condition $\cos(n \frac{\pi L}{\ell} + \tilde{\delta}_n L) = 1$ reads $n \frac{\pi L}{\ell} + \tilde{\delta}_n L = m 2 \pi$, with $m = 0, 1, 2, \dots$, or $\tilde{\delta}_n = m 2 \pi/L - n \pi/\ell$. Note that $\tilde{\delta}_0 \in I_0$ if and only if $m=0$, since $\ell>L$, and thus $\tilde{\delta}_0 = 0$. For $\tilde{\delta}_n$ to belong to $I$, with $n\geq 1$, it is necessary that $|4 m \ell/L - 2n| \leq 1$. Let us assume that $\ell/L$ is an integer (as in the simulation), then only the condition $|4 m \ell/L - 2n| = 0$ or $\tilde{\delta}_n=0$ might be satisfied, and thus any exceptional point $\tilde{\delta}_n$ must be at the centre of $I$. Let us introduce the set $\Sigma = \{2 m \ell/L \ | \ m=1,2,\dots \}$, which is the collection of $n$ for which the condition $|4 m \ell/L - 2n| = 0$, with $m=1,2,\dots$, is satisfied. If $n \in \Sigma$, then we shall cut the interval $I$ into two parts: $I = I^{\rm L} \cup I^{\rm R}$, where $I^{\rm L} = [-\frac{\pi}{2\ell},0]$ and $I^{\rm R}=[0,\frac{\pi}{2\ell}]$.

Let us define $F_0 : I_0 \rightarrow I_0$ as
\begin{equation}\label{Equ F0}
F_0(\delta) = \frac{1}{\ell} \arctan\left[\frac{\sin(\delta L)}{2 [1-\cos(\delta L)]} \right]~,
\end{equation}
and set $F_0(0)=\pi/(2 \ell)$. If $n \geq 1$ and $n \not \in \Sigma$, then we define $F_n : I \rightarrow I$ by
\begin{equation}\label{Equ F}
F_n(\delta) = \frac{1}{\ell} \arctan\left[\frac{\sin(n \frac{\pi L}{\ell} + \delta L)}{2 [1-\cos(n \frac{\pi L}{\ell} + \delta L)]} \right]~.
\end{equation}
On the other hand, if $n \in \Sigma$, we shall consider the functions $F^{\rm L} : I^{\rm L} \rightarrow I^{\rm L}$ and $F^{\rm R} : I^{\rm R} \rightarrow I^{\rm R}$ given by 
\begin{eqnarray}
F^{\rm L}(\delta) &=& \frac{1}{\ell} \arctan\left[\frac{\sin(\delta L)}{2 [1-\cos(\delta L)]} \right]~,\\
F^{\rm R}(\delta) &=& \frac{1}{\ell} \arctan\left[\frac{\sin(\delta L)}{2 [1-\cos(\delta L)]} \right]~,
\end{eqnarray}
and set $F^{\rm L}(0) = -\pi/(2 \ell)$ and $F^{\rm R}(0) = \pi/(2 \ell)$. 

We shall show that $F_0$, $F_n$, $F^{\rm L}$ and $F^{\rm R}$ are contracting maps and thus, by the Fixed Point Theorem, there exist unique points $\delta_0^*$, $\delta_n^*$, $\delta^{\rm L *}$ and $\delta^{\rm R *}$  such that $F_0(\delta_0^*)=\delta_0^*$, $F_n(\delta_n^*)=\delta_n^*$, $F^{\rm L}(\delta^{\rm L *})=\delta^{\rm L *}$ and $F^{\rm R}(\delta^{\rm R *})=\delta^{\rm R *}$. Furthermore, these fixed points may be reached by iterations, \eg $F^m_n(0)$, the $m$-th iterate of $F_n$ at point $0$, converges to $\delta_n^*$ as $m \rightarrow \infty$. This shows in particular that we have found all the solutions of \eref{Equ Cn} and consequently of \eref{Implicit}, that $H_0$ has pure point spectrum and that each eigenvalue has multiplicity 1. Let us now show that $F_n$ contracts, the proofs for $F_0$, $F^{\rm L}$ and $F^{\rm R}$ are similar. So let $X, Y \in I$ with $X \not = Y$. Then,
\begin{equation}
|F_n(X)-F_n(Y)| \leq \mathcal{C} \ |X-Y|~,
\end{equation}
where
\begin{equation}
 \mathcal{C} = \frac{|F_n(X)-F_n(Y)|}{|X-Y|} \leq \sup_{X\not=Y \in I} \frac{|F_n(X)-F_n(Y)|}{|X-Y|} \leq \sup_{X \in I} \left|\frac{dF_n}{dX}(X)\right| \leq \frac{L}{\ell} < 1~.
\end{equation}
The last two inequalities follow from the assumption $L < \ell$ and the explicit computation of $F'(X)$:
\begin{equation}
\left|\frac{dF_n}{dX}(X)\right| = \frac{2L}{\ell [5 - 3 \cos(n \frac{\pi L}{\ell} + LX)]}~.
\end{equation}
It remains to check that the fixed point $\delta_n^*$ of $F_n$ gives rise to an allowed value of $k_n$, \ie satisfying  $k_n \not \in \{2 \pi m/L, (2m+1) \pi/(2\ell)\}_{m=0}^{\infty}$. 
These conditions state that $\delta_n^*$ should not be a boundary value of $I$, \ie $|\delta_n^*| \not = \pi/(2\ell)$. This can be easily checked.

In the simulations, we iterate the map $F_0$, $F_n$, $F^{\rm L}$ and $F^{\rm R}$ until the condition \eref{Condition 3p} with $\psi=\varphi^{\rm global}_n$ given in \eref{Global sol} is accurately satisfied: we required that $|\varphi^{\rm global}_n(L+\ell)| < 10^{-10}$. Since the contracting constant $\mathcal{C}$ is very small ($<L/\ell$), this condition is satisfied after a only few iterations. In summary, we have
\begin{eqnarray}
 k_0 &=& \delta_0^*~, \mbox{ with } \delta_0^* \approx F_0\circ \cdots \circ F_0(\epsilon)~,\\
n \not \in \Sigma: k_n &=& n \frac{\pi}{\ell} + \delta_n^*~, \mbox{ with } \delta_n^* \approx F_n\circ \cdots \circ F_n(0)~,\\
n  \in \Sigma: k^{\rm L}_n &=& n \frac{\pi}{\ell} + \delta^{\rm L *}~, \mbox{ with } \delta^{\rm L *} \approx F^{\rm L}\circ \cdots \circ F^{\rm L}(-\epsilon)~,\\
n  \in \Sigma: k^{\rm R}_n &=& n \frac{\pi}{\ell} + \delta^{\rm R *}~, \mbox{ with } \delta^{\rm R *} \approx F^{\rm R}\circ \cdots \circ F^{\rm R}(\epsilon)~,
\end{eqnarray}
where $\epsilon$ is a small positive number [$\epsilon < \pi/(2\ell)$]. The states can then be written as
\begin{eqnarray}
\begin{matrix}
\mbox{ if } n \not \in \Sigma: &  \varphi_n = \varphi^{\rm global}_n~,\\ \\
\mbox{ if } n \in \Sigma: &
\begin{cases} \varphi^{\rm L}_n = \varphi^{\rm global}_n, & \mbox{ with } \hspace{3mm} k_n=k^{\rm L}_n\\
                      \varphi^{\rm C}_n = \varphi^{\rm loop}_n, & \mbox{ with } \hspace{3mm} k_n=\frac{n\pi}{\ell}\\
                      \varphi^{\rm R}_n = \varphi^{\rm global}_n, & \mbox{ with } \hspace{3mm} k_n=k^{\rm R}_n 
\end{cases}
\end{matrix}
\end{eqnarray}


\begin{thebibliography}{28}
\bibitem[{Com({\natexlab{a}})}]{CommentS}
\bibinfo{note}{In some papers discussing only $P(t)$, the name "survival
  probability" is used instead of "nonescape probability".}
\bibitem[{Com({\natexlab{b}})}]{Comment1}
\bibinfo{note}{We have checked that the same results are obtained if one
  computes the scattering states $\varphi_k$ in the local reference frame (see
  \Fig 2) by using the scattering matrix $S$.}
\bibitem[{Bou()}]{Bound}
\bibinfo{note}{If $\psi_0 \in D(H)$ is a normalized bound state of $H$, \ie $H
  \psi_0 = E \psi_0$ with $E > 0$ and $\int_0^L |\psi_0(x)|^2 dx = 1$, then
  $\psi(x,t) = (e^{-iHt}\psi_0)(x) = e^{-iEt}\psi_0(x)$ so that $P(t) =
  \int_0^L |\psi(x,t)|^2 dx = \int_0^L |\psi_0(x)|^2 dx = 1$ for all times.
  This argument can then be extended to any linear combination of bound states
  with different eigenvalues.}
\end{thebibliography}

\end{document}